\documentclass[trackchanges,twocolumn]{aastex701}

\defcitealias{WeissermanGillis2025}{W25}

\begin{document}

\title{New and Updated Rossiter-McLaughlin Measurements for Three Hot Jupiter-Hosting M Dwarfs}

\author[0000-0002-7992-469X]{Drew Weisserman}
\affiliation{Department of Physics \& Astronomy, McMaster University, 1280 Main St W, Hamilton, ON, L8S 4L8, Canada}
\email[show]{weisserd@mcmaster.ca}

\author[0000-0001-5383-9393]{Ryan Cloutier}
\affiliation{Department of Physics \& Astronomy, McMaster University, 1280 Main St W, Hamilton, ON, L8S 4L8, Canada}
\email{ryan.cloutier@mcmaster.ca}

\author[0009-0008-1229-3230]{Alexandra Rochon}
\affiliation{Department of Physics \& Astronomy, McMaster University, 1280 Main St W, Hamilton, ON, L8S 4L8, Canada}
\email{rochoa3@mcmaster.ca}

\author[0009-0003-2931-9525]{Ze'ev Vladimir}
\affiliation{Department of Astronomy \& Astrophysics, University of Chicago, Chicago, IL 60637, USA}
\email{zvladimir@arizona.edu}

\author[0000-0002-8868-7649]{Bertram Bitsch}
\affiliation{Department of Physics, University College Cork, College Rd, Cork T12 K8AF, Ireland}
\email{bbitsch@ucc.ie}

\author[0000-0003-4508-2436]{Ritvik Basant}
\affiliation{Department of Astronomy \& Astrophysics, University of Chicago, Chicago, IL 60637, USA}
\email{rbasant@uchicago.edu}

\author[0000-0003-4733-6532]{Jacob L. Bean}
\affiliation{Department of Astronomy \& Astrophysics, University of Chicago, Chicago, IL 60637, USA}
\email{jacobbean@uchicago.edu}

\author[0009-0005-1486-8374]{Tanya Das}
\affiliation{Department of Astronomy \& Astrophysics, University of Chicago, Chicago, IL 60637, USA}
\email{tanyadas@uchicago.edu}

\author[0000-0001-9796-2158]{Emily Deibert}
\affiliation{Department of Physics \& Astronomy, University of Waterloo, 200 University Avenue West, Waterloo, ON, N2L 3G1, Canada}
\email{emily.deibert@uwaterloo.ca}

\author[0000-0001-5442-1300]{Thomas M. Evans-Soma}
\affiliation{School of Science, University of Newcastle, Callaghan, NSW, Australia}
\email{tom.evans-soma@newcastle.edu.au}

\author[0000-0002-2338-476X]{Michael Line}
\affiliation{School of Earth and Space Exploration, Arizona State University, Tempe, AZ 85281, USA}
\email{mrline@asu.edu}

\author[0000-0002-7605-2961]{Ralph Pudritz}
\affiliation{Department of Physics \& Astronomy, McMaster University, 1280 Main St W, Hamilton, ON, L8S 4L8, Canada}
\email{pudritz@mcmaster.ca}

\author[0000-0003-4526-3747]{Andreas Seifahrt}
\affiliation{Gemini Observatory/NSF NOIRLab, 670 N. A'ohoku Place, Hilo, HI 96720, USA}
\email{andreas.seifahrt@noirlab.edu}

\author[0000-0002-7260-5821]{Evgenya L. Shkolnik}
\affiliation{School of Earth and Space Exploration, Arizona State University, Tempe, AZ 85281, USA}
\email{shkolnik@asu.edu}

\author[0000-0003-0156-4564]{Luis Welbanks}
\affiliation{School of Earth and Space Exploration, Arizona State University, Tempe, AZ 85281, USA}
\email{luis.welbanks@asu.edu}

\begin{abstract}
Evidence suggests that high-eccentricity migration (HEM) is the dominant migration channel for short-period Giant Exoplanets around M dwarf Stars (GEMS). However, it is unlikely that all short-period GEMS form via HEM, given that most systems lack known massive companions capable of driving HEM. We present Rossiteer-McLaughlin (RM) effect detections for the GEMS TOI-5205 b, TIC 46432937 b, and TOI-3714 b using the Gemini-North/MAROON-X spectrograph, bringing the total number of GEMS with RM detections to five. Our systems are well-aligned, with sky-projected obliquities of {$\lambda = 0_{-5}^{+6}$$^\circ$, $0 \pm 3^\circ$, and $12_{-7}^{+9}$$^\circ$}, respectively, and we measure a deprojected obliquity of {$\psi = 29 \pm 10^\circ$} for TOI-3714. We analyze archival radial velocities, astrometry, and speckle imaging data to search for additional companions around all five known GEMS with RM detections. We find tentative evidence for a new massive companion around one of these GEMS, TOI-5293 A, in Gaia DR2+DR3 data, though further follow-up is needed for confirmation. We rule out massive companions between $\sim 1-10{\rm  AU}$ in the remaining systems, but cannot rule out all companions capable of driving HEM. Our findings present further evidence that short-period GEMS are preferentially aligned. {While these GEMS feasibly have sufficient orbital angular momentum to realign their host stars, tidal realignment may take implausibly long; either alternative explanations for fast tidal realignment of GEMS or reconciliations of disk-driven migration with circumstantial evidence suggestive of HEM driving GEMS migration must be found.}

\end{abstract}

\section{Introduction}\label{sec:intro}



Short-period Giant Exoplanets around M dwarf Stars (GEMS) are $2-3\times$ less common than hot Jupiters (HJs) around AFGK stars \citep{JohnsonAller2010,BeleznayKunimoto2022,BryantBayliss2023,GanWang2023}. This discrepancy is corroborated by models of gas giant formation via core accretion, which struggle to produce GEMS due to the low solid disk masses around their would-be stellar hosts \citep{LaughlinBodenheimer2004,MoralesMustill2019}, and suggests that giant planet formation is highly sensitive to the host star and its natal protoplanetary disk. By extension, hot Jupiters likely form and migrate via distinct pathways around M dwarfs compared to around AFGK stars. However, due to the comparatively low occurrence rate of GEMS  \citep{BryantBayliss2023,GanWang2023} and the faintness of their host stars, these planets remain relatively understudied. 
Fortunately, NASA’s Transiting Exoplanet Survey Satellite \citep[TESS;][]{RickerWinn2015} has substantially expanded the sample of confirmed GEMS, many of which are amenable to detailed follow-up characterization efforts necessary for tracing these planets' formation and migration processes.


\defcitealias{FrenschBouchy2026}{Y. Frensch et al. 2026} 
Evidence is mounting for a form of high-eccentricity migration (HEM) plus tidal damping as the dominant formation mechanism of HJs around AFGK stars. Nearby planetary systems containing HJs tend to lack nearby small planets \citep{LathamRowe2011,HuangWu2016}, potentially indicative of planetary system disruption during the epochs of dynamically hot HEM. {Systems with HJs are also more likely to have distant binary companions than field stars are \citep{NgoKnutson2015}}. In addition, those binary companions are preferentially misaligned with the HJs relative to other types of transiting planets, as would be expected from Kozai-Lidov (KL) migration \citep{BehmardDai2022,ChristianVanderburg2025}. Furthermore, analyses of the HJ eccentricity-obliquity-$T_{\rm eff}$ distributions \citep{RiceWang2022} and age distributions \citep{SchmidtSchlaufman2026} of HJs show that both are consistent with HEM plus tidal damping as the dominant migration mechanism of these planets. 

However, probing the dynamical histories of GEMS remains an active area of study. \citet{WeissermanGillis2025} (hereafter \citetalias{WeissermanGillis2025}) calculated the HJ formation efficiency via the KL mechanism from the occurrence of HJ progenitors (i.e. cold Jupiters) and the multiple fraction of misaligned stellar companions capable for driving KL migration. Those calculations revealed that the HJ formation efficiency by KL is higher for GEMS than for HJs around AFGK stars \citepalias{WeissermanGillis2025}, suggesting that HEM may be the dominant migration mechanisms for GEMS as well. It remains improbable that all GEMS formed via KL migration. While the stellar multiplicity of M dwarfs hosting HJs is higher than in the field \citep[\citetalias{WeissermanGillis2025};][]{FrenschBouchy2026,GanLHeureux2026,MessamahBouchy2026}, many confirmed close-in GEMS show no evidence for the widely separated, misaligned, massive companions that KL migration requires. KL migration may be the dominant migration channel for close-in GEMS, but the massive companions needed to drive KL migration do not appear to be strictly required for them to form. 


Measurements of the Rossiter-McLaughlin (RM) effect \citep{Rossiter1924,McLaughlin1924} can be used to characterize stellar obliquities, which probe formation and evolutionary histories of transiting planets. To date, however, the RM effect has only been successfully detected for three GEMS: TOI-4201 b \citep{GanWang2024}, TOI-3714 b, and TOI-5293 b \citepalias{WeissermanGillis2025}. All three of these GEMS systems are consistent with having well-aligned stellar obliquities. This finding is simultaneously consistent with {disk-driven migration that preserves the obliquity from a primordially aligned disk} and with a history of HEM followed by efficient tidal damping. Cool stars in particular are thought to allow for rapid tidal dissipation that would efficiently dampen misaligned obliquities excited by HEM \citep{WinnFabrycky2010,Attia2023,SpaldingWinn2022}. 

In this work, we present new detections of the RM effect for two short-period GEMS, TOI-5205 b and TIC 46432937 b. We also present a second transit observation of TOI-3714 b and refine its obliquity, previously measured in \citetalias{WeissermanGillis2025}. Our results bring the sample of short-period GEMS with RM detections from three to five. Our paper is structured as follows: in Section \ref{sec:data}, we describe our observations. In Section \ref{sec:analysis}, we describe our data analysis methodology and present the results of our RM effect measurements. We discuss the implications of our results on these planets' formation histories in Section \ref{sec:disc}, and conclude with a summary of our findings in Section \ref{sec:conc}.

\section{Data}\label{sec:data}

We observed six full or partial transit sequences, alongside pre- and post-transit baselines, of our three short-period GEMS targets. We observed all transits with the MAROON-X spectrograph, a high-resolution ($R\sim 85,000)$ optical \'{e}chelle spectrograph located at the 8m Gemini-North Telescope on Maunakea, Hawai'i \citep{SeifahrtSturmer2018,SeifahrtBean2020}. Our observations were taken in the spectrograph's blue ($500-670$ nm) and red ($650-900$ nm) arms simultaneously. All exposure times presented in this work were fixed to 900 seconds. One out of two transit sequences of TOI-3714 was obtained under the queue program GN-2024B-Q-132 (PI: Cloutier) and was originally published in \citetalias{WeissermanGillis2025}. The five remaining and previously unpublished transit sequences were obtained under GN-2025B-Q-226 (PI: Cloutier). All our raw MAROON-X data were reduced using custom routines originally written for VLT/CRIRES \citep{BeanSeifahrt2010}. We extracted radial velocity (RV) measurements from both the blue and red arms using the \texttt{SERVAL} template-matching code \citep[SpEctrum Radial Velocity AnaLyzer;][]{ZechmeisterReiners2018}, and combined them into a single RV estimate per exposure {by calculating a weighted average of the two exposures, weighted according to the inverse square of the RV uncertainty for each arm}. All RVs from RM measurements modeled in this work are available in Table \ref{tab:ap-rvs}.

\subsection{TOI-5205}\label{subsec:data-TOI5205}

\citet{KanodiaMahadevan2023} reported the discovery of a hot Jupiter orbiting around an M4V star in the TOI-5205 system. This planet, TOI-5205 b, orbits at a period of $1.63$ days and has a mass of $1.08 \pm 0.06 M_J$ and a radius of $1.03 \pm 0.03 R_J$ \citep{KanodiaMahadevan2023}.

This planet's host star, TOI-5205, is not known to have a stellar companion. This star has a mass of $0.392 \pm 0.015 M_\odot$ and a K-band magnitude of $11.04 \pm 0.02$ \citep{KanodiaMahadevan2023}. TOI-5205 does not have a confirmed rotation period; \citet{KanodiaMahadevan2023} does not identify a variability in TESS light curves of this star and cannot constrain it from rotational broadening of HPF spectra.

We observed three transits of TOI-5205 b (denoted Transits 1, 2, and 3) on August 2, 2025, August 20, 2025, and September 2, 2025 UT. Transit 1 observed a partial transit, with {six} exposures spanning airmasses between 1.00 and 1.05. This transit was taken in poor observing conditions due to fog, and the seeing substantially degraded over the course of the observation; as such, the observation was called off partway through the transit, spanning the pre-ingress baseline and roughly half of the transit duration. Reduction of the observations taken in the blue arm failed entirely for this transit, so the reduction for just the red arm is used as a result. Transit 2 observed a full transit, with thirteen exposures spanning airmasses between 1.00 and 1.14. Transit 3 observed a full transit, with twelve exposures spanning airmasses between 1.02 and 1.68. Our RV extractions yielded RV uncertainties in the blue channel of $8.1$ and $6.7$ m/s for Transits 2 and 3 respectively; median RV uncertainties in the red channel of $7.0$, $3.3$, and $2.7$ m/s for Transits 1, 2, and 3 respectively; and median combined RV uncertainties of $7.0$, $3.1$, and $2.5$ m/s for Transits 1, 2, and 3 respectively.

\subsection{TIC 46432937}\label{subsec:data-tic46432937}
\citet{HartmanBayliss2024} reported the discovery of a hot Jupiter orbiting the early-M dwarf TIC 46432937. This planet, TIC 46432937 b, orbits at a period of $1.44$ days and has a mass of $3.20 \pm 0.11 M_J$  and a radius of $1.188 \pm 0.030 R_J$ \citep{HartmanBayliss2024}.

TIC 46432937 {has a distant co-moving companion at a projected distance of $\sim 3600$ AU \citep{GanLHeureux2026}}. This star has a mass of $0.563 \pm 0.029 M_\odot$ and a K-band magnitude of $10.195 \pm 0.020$ \citep{HartmanBayliss2024}. TIC 46432937 does not have a confirmed rotation period. \citet{HartmanBayliss2024} identifies possible variability in this star's TESS light curves at a period of $5.88 \pm 0.54$ days; however, this study is unable to rule out uncorrected instrumental effects, and so is unable to confirm any rotation period for this star.

We observed one full transit of TIC 46432937 b on December 27, 2025 UT. This transit consisted of eleven exposures, spanning airmasses between 1.21 and 1.61. The RV extractions yielded a median RV uncertainty in the blue channel of $3.5$ m/s, a median RV uncertainty in the red channel of $2.7$ m/s median red RV uncertainty of $2.7$ m/s, and a median combined RV uncertainty of $3.0$ m/s.

\subsection{TOI-3714}\label{subsec:data-TOI3714}

\citet{CanasKanodia2022} reported the discovery of a hot Jupiter on an S-type orbit around the M2 dwarf companion star in the TOI-3714 binary system. This planet orbits at a short period of $2.15$ days \citep{CanasKanodia2022} and has a mass of $0.67 \pm 0.02 M_J$ \citepalias{WeissermanGillis2025} and a radius of $1.01 \pm 0.03 R_J$ \citep{CanasKanodia2022}.

The primary star in the system is a white dwarf with a mass of $\sim 1.07 M_\odot$ and an estimated cooling age of $2.4$ Gyr \citep{CanasKanodia2022}. The planet-hosting companion star, TOI-3714, has a mass of $0.507 \pm 0.011 M_\odot$ \citepalias{WeissermanGillis2025} and a $K$-band magnitude of $10.852\pm 0.017$ \citep{CanasKanodia2022}. TOI-3714 does have a measured rotation period of $23.3 \pm 0.3$ days \citep{CanasKanodia2022}; knowing this star's rotation period can allow for measurements of the star's deprojected stellar obliquity $\psi$ (which is done in Section \ref{subsec:disc-deproj}).

\citetalias{WeissermanGillis2025} presented one full transit of TOI-3714 b, Transit 1, on December 10, 2024 UT, which consisted of sixteen exposures and spanned airmasses between 1.06 and 1.24. This paper additionally presents a full transit observed on September 16, 2025 UT, Transit 2, which consisted of thirteen exposures and spanned airmasses between 1.09 and 1.90. These RV extractions yielded median RV uncertainties in the blue channel of $8.6$ and $4.7$ m/s, median RV uncertainties in the red channel of $5.6$ and $3.1$ m/s, and median combined RV uncertainties of $4.7$ and $3.6$ m/s for Transits 1 and 2 respectively.

\section{Data Analysis \& Results}\label{sec:analysis}




\subsection{RV Model} \label{subsec:analysis-rvmodel}

We use the \texttt{starry} package \citep{LugerAgol2019} to model both Keplerian RV signals and the RM effect. Within \texttt{starry}, the RM effect is modeled by a star whose surface brightness is uniform, as parametrized by the spherical harmonic with degree and order $\{l, m\} = \{0, 0\}$, but also includes quadratic limb darkening. Due to the overlapping wavelength coverage of TESS and MAROON-X, we adopt the TESS-band limb-darkening parameters from each planet's discovery paper. (For TOI-5205, this is $u_1, u_2 = 0.61, -0.12$, as measured in TESS sector 41, the most recent TESS sector modeled in \citet[][priv. comm.]{KanodiaMahadevan2023}, while for TOI-3714, we use the TESS limb-darkening parameters from \citealt{CanasKanodia2022}). {While the wavelength ranges between TESS and MAROON-X are not identical, the wavelengths do overlap, justifying this choice; furthermore, even unphysical limb darkening parameters will only negligibly affect the resulting obliquity posterior, and so our results are robust to the limb darkening parameters used.}

Our full RV model features {seven} parameters, plus an additional {one} for each transit. The Keplerian orbit is parameterized by the planet's orbital period $P$, time of mid-transit $t_0$, and the RV semi-amplitude $K$. {We assume $e = 0$ for each planet, as they are expected to have rapidly tidally circularized.} The RM effect is parameterized by the planet-to-star radius ratio, $R_p/R_\star$, the transit inclination $i$ (or equivalently, the transit impact parameter $b$), the projected stellar rotation velocity $v \sin i_*$, and the projected stellar obliquity $\lambda$. In addition, we fit an RV offset $v_{0,n}$ for each transit number $n$.

We note that our model ignores the effects of stellar activity. M dwarf activity at field ages is dominated by rotationally-modulated magnetically active regions. While active regions occulted by the planet during the transit could produce an observable signature, we cannot reliably assess the occurrence of such an occulted region with our MAROON-X data. We note that we expect the impact of occulted starspots and plages to be negligible, however -- the photometric and spectroscopic youth indicators from TOI-5205 b's discovery paper suggests that TOI-5205 is a low-activity, old star \citep{KanodiaMahadevan2023}, while a lack of photometric variability attributable to stellar variability suggests the same for TOI-3714 and TIC 46432937 \citep{CanasKanodia2022,HartmanBayliss2024}.

Our program seeks observing baselines that span two transit durations, each of which is less than two hours. The short durations of our observing baselines with MAROON-X provide very weak independent constraints on the Keplerian and transit parameters compared to the inferences from each planet's transit and orbital RV time series. We therefore need not reproduce those analyses and instead adopt the parameter posteriors for $\{M_\star, R_\star, P, t_0, K, i, R_p/R_\star, e\}$ from those studies \citep[i.e.][]{CanasKanodia2022,KanodiaMahadevan2023,HartmanBayliss2024,WeissermanGillis2025} as priors. The MAROON-X data presented in this paper therefore provides novel constraints on the parameters $\{v_{0}, v\sin{i_\star}, \lambda\}$. Our model parameter priors are reported in Section \ref{ap:posteriors}.

Due to the sharp inflections in the RM model that occur on time scales $\lesssim t_{\mathrm{exp}}$ (i.e. shortly after transit ingress and shortly before transit egress), caution must be taken to compare the observed RVs to the time-averaged model RV across a given observation rather than to the model RV at the midpoint of the observation. As such, when evaluating the model to compare to observations, we oversample the model by a factor of 20 over the timespan of the observations. We then compute the average of the model across the exposure time of each observation to calculate the likelihood function, averaging together 20 evenly spaced sample points across each 900 second exposure.

We use the \texttt{emcee} Markov chain Monte Carlo (MCMC) package \citep{ForemanMackeyHogg2013} to sample the posterior of our full RV model for each planet. We use 64 walkers to sample parameter space for each target, checking every 20,000 steps until our chain length corresponds to $\sim 30\times$ the autocorrelation time. This took 180,000, 140,000, and 80,000 steps for TOI-5205, TIC 46432937, and TOI-3714, respectively. We then omit the first 20\% of steps as burn-in. Our point estimates of the median, 16th, and 84th percentiles of the resulting marginalized posterior of each parameter are reported in Tables \ref{tab:TOI5205-posterior}, \ref{tab:tic46432937-posterior}, and \ref{tab:TOI3714-posterior}.

\begin{figure}[!h]
    \centering
    \includegraphics[width=.9\linewidth]{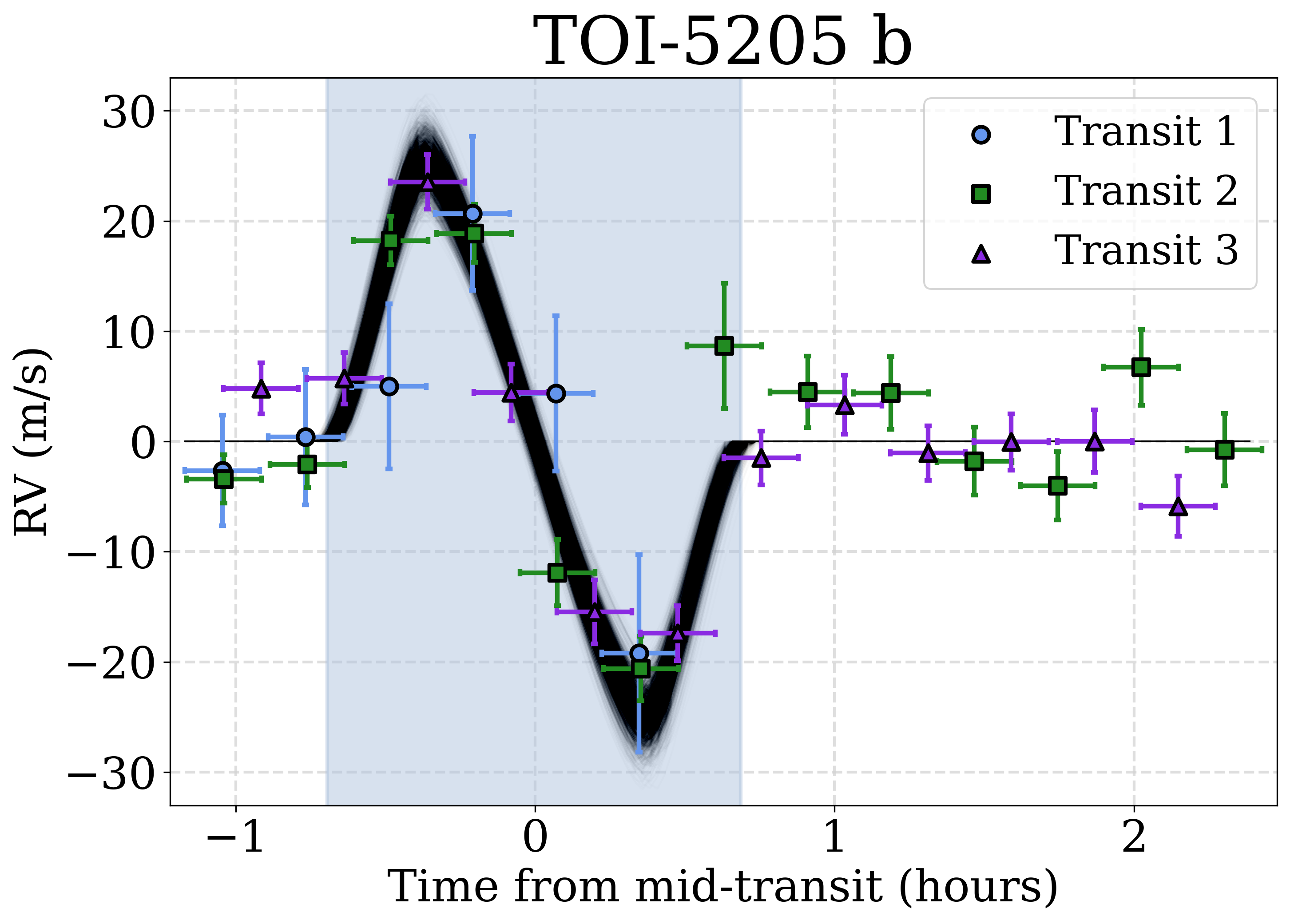}
    \centering
    \includegraphics[width=.9\linewidth]{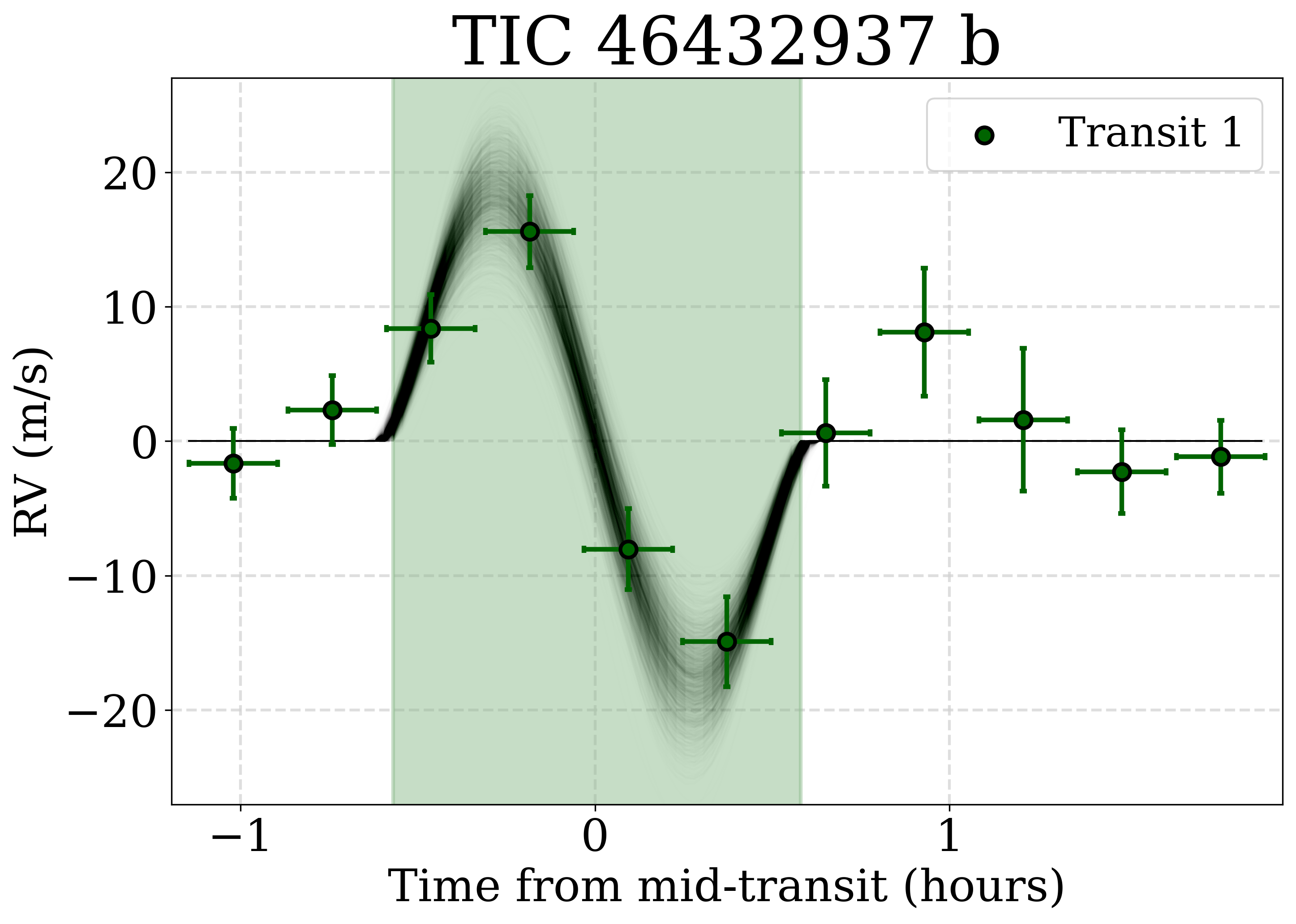}  
    \centering
    \includegraphics[width=.9\linewidth]{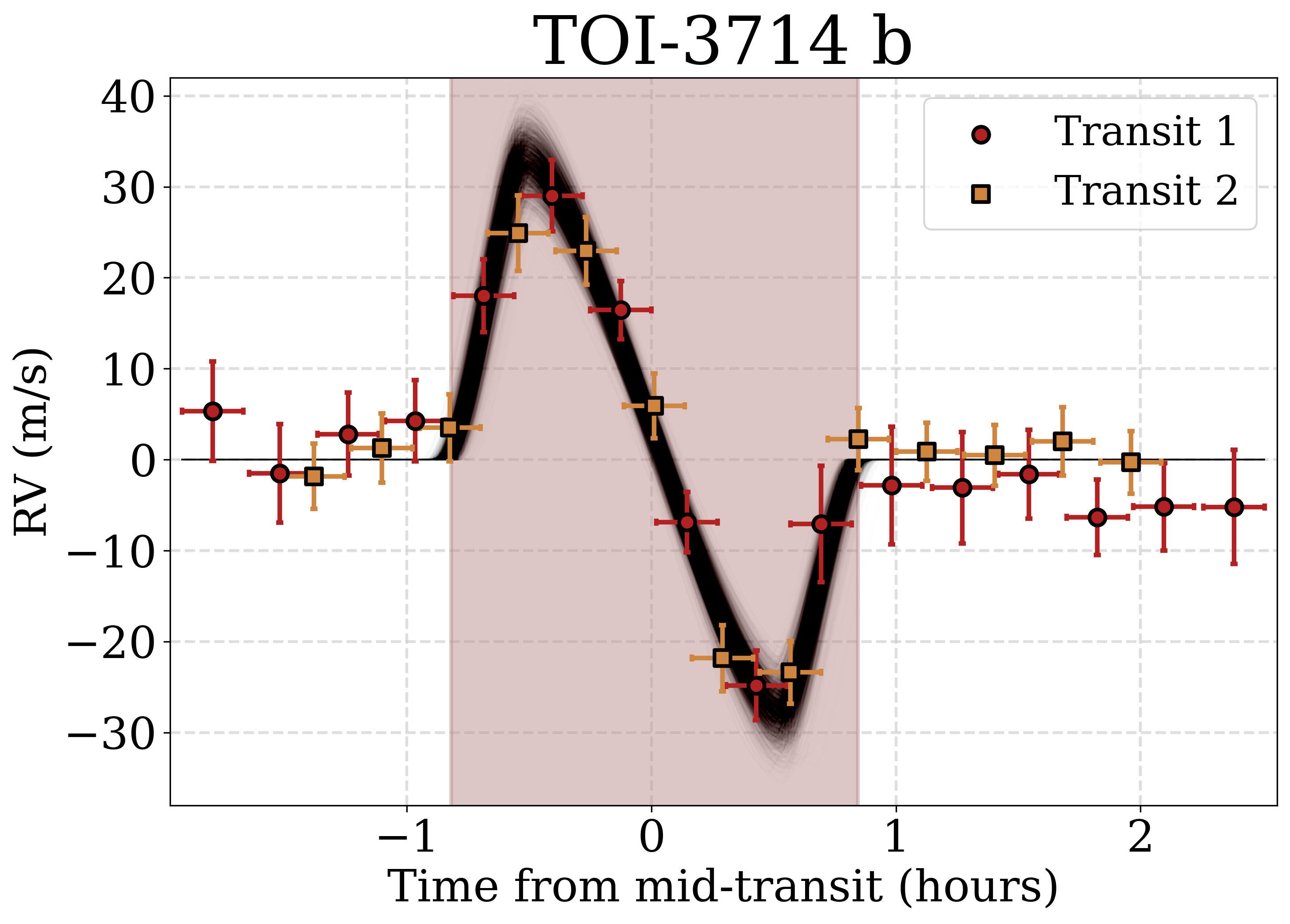}
    \caption{Detections of the RM effect for our three hot Jupiters orbiting M dwarfs. The black curves represent random draws from our RM model posteriors after removing the median detrending model for each transit and the Keplerian orbital signals. The duration of the planetary transit is shaded.}
        \label{fig:rmplots}
\end{figure}

Figure \ref{fig:rmplots} depicts the results of our RV modeling for all three targets. We detect the RM effect for all three planets at high significance, with semi-amplitudes of {$24.1_{-1.6}^{+1.5}$ m/s ($16\sigma$), $25.8_{-3.1}^{+3.3}$ m/s ($8.2\sigma$), and $28.7_{-1.8}^{+2.1}$ m/s ($16\sigma$)} for TOI-5205 b, TIC 46432937 b, and TOI-3714 b, respectively. We measure sky-projected stellar obliquities of {$\lambda = 0_{-5}^{+6}$$^\circ$, $0\pm3^\circ$, and $12_{-7}^{+9}$$^\circ$}, and projected stellar rotation velocities of {$v\sin {i_\star} = 0.53 \pm 0.04$ km/s, $1.32 _{-0.18}^{+0.20}$ km/s, and $1.09_{-0.09}^{+0.10}$} km/s for TOI-5205, TIC 46432937 b, and TOI-3714, respectively. Our measured $\lambda$ and $v\sin{i_\star}$ values for TOI-3714 are consistent within $1\sigma$ of previously measured values from \citetalias{WeissermanGillis2025} ($\lambda=21_{-11}^{+14}$$^\circ, v\sin{i_\star} =1.06_{-0.12}^{+0.15}$ km/s), with our revised measurements being more precise by roughly the amount expected from coadding two transit observations of similar quality (i.e. RV uncertainty).

\section{Discussion}\label{sec:disc}

\begin{figure*}
    \centering \includegraphics[width=0.9\linewidth]{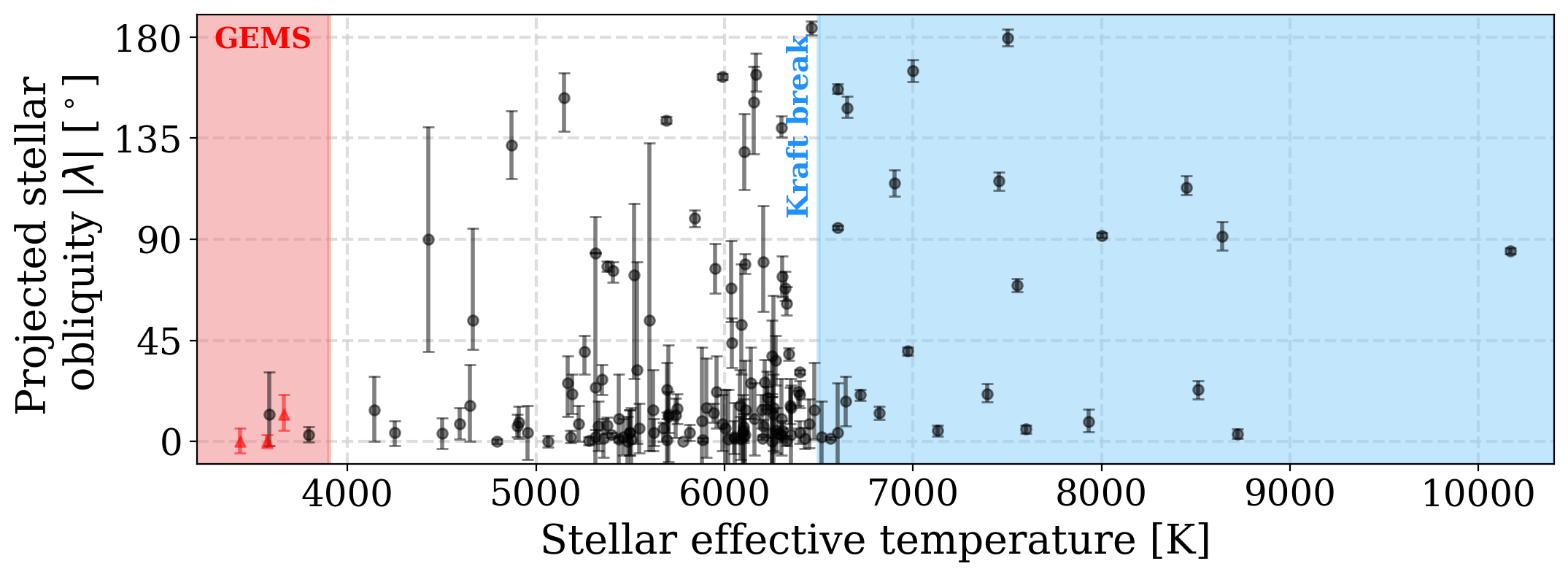}
    \caption{The distribution of projected stellar obliquities $|\lambda|$ for giant exoplanets ($R_p > 8R_\oplus$) as a function of host star effective temperature. All stellar obliquity measurements are retrieved from the NASA Exoplanet Archive, retrieved on June 30, 2026. Short-period GEMS occupy the shaded red region of parameter space with $T_{\rm eff} < 3900 {\rm  K}$, while host stars above the Kraft break {\citep[$T_{\rm eff} > 6500 {\rm  K}$;][]{WangWang2026}} occupy the shaded blue region. The three short-period GEMS whose obliquities are measured in this work are denoted with red triangles.}
    \label{fig:oblplot}
\end{figure*}

Our detections of the RM effect for TOI-5205 b and TIC 46432927 b represent the fourth and fifth such detections for short-period GEMS. Figure~\ref{fig:oblplot} compares our projected stellar obliquities of gas giant hosts as a function of host star effective temperature. The projected stellar obliquities of all three GEMS measured in our work -- and all five short-period GEMS with RM measurements in the literature -- are consistent with being aligned.


\subsection{Deprojected Stellar Obliquities}\label{subsec:disc-deproj}

While the RM effect is sensitive to the sky-projected stellar obliquity $\lambda$, detailed inferences on the dynamical histories of individual transiting planets require the deprojected stellar obliquity $\psi$. The deprojected stellar obliquity is calculated from

\begin{equation}
\cos \psi = \cos i_\star \cos i_p + \sin i_\star \sin i_p \cos \lambda,
\end{equation}

\noindent where $i_*$ is the inclination angle of the host star’s rotation, $i_p$ is the orbital inclination of the transiting planet, and $\lambda$ is the projected stellar obliquity from the RM effect \citep{FabryckyWinn2009}. 

$i_\star$ can be recovered from knowledge of $v \sin i_\star$ and $P_{rot}$ alongside the stellar radius. Unfortunately, while TOI-3714 has a measured rotation period of $23.3 \pm 0.3$ days \citep{CanasKanodia2022}, TIC 46432937 does not have a known rotation period, and so we cannot measure $i_\star$ nor $\psi$ for this target. It should be noted that \citet{HartmanBayliss2024} presented a candidate rotation period for TIC 46432937 of $5.88 \pm 0.54$ days, but was unable to rule out uncorrected instrumental effects \citep{HartmanBayliss2024}. We note here as well that the TESS All-Sky Rotation Survey \citep[TARS;][]{BoyleBouma2026} did attempt to measure a rotation period for TIC 46432937. TARS performs additional tests to identify true stellar variability and distinguish it from systematics, including training a classifier to identify systematic signals and checking for contamination from neighboring stars; after these tests, TARS does not identify a rotation period for TIC 46432937. As such, we do not consider the 5.88 day possible rotation period to be a true measure of the stellar rotation period, and so cannot properly characterize TIC 46432937's inclination or deprojected obliquity.

TOI-5205 does have a measured rotation period of $22.8\pm 2.9$ days from TARS. However, we caution against the use of this rotation period for TOI-5205. In the four sectors of TESS data that TARS analyzes for TOI-5205 (Sectors 15, 41, 55, and 82), it only classifies the light curve as containing periodic variability in two of them (41 and 55). In addition, visual inspection of the light curves reveals a signal that does not look like rotation, either at $22.8$ days or at the half-period harmonic of $11.4$ days that TARS resolves. \citet{BoyleBouma2026} estimates that $93\%$ of the periods reported in TARS are rotation periods; however, we suspect any periodicity for TOI-5205 is erroneous or mischaracterized. As such, we only characterize the deprojected obliquity and inclination of TOI-3714 here. 

We recover the values of $i_\star$ for TOI-3714 using a-priori knowledge of their stellar radii and photometric rotation periods along with our measurements of $v\sin i_\star$ from the RM effect. We follow the MCMC-based methodology of \citet{MasudaWinn2020} to probabilistically infer $\cos i_\star$ from independent datasets that measure $v_{eq} = 2\pi R_\star/P_{\rm rot}$ and $v \sin i_\star$, respectively.

We derive {$i_\star = 64_{-11}^{+12}$$^\circ$} for TOI-3714, yielding a deprojected stellar obliquity of {$\psi = 29 \pm 10^\circ$}. This value is consistent with the deprojected obliquity of $\psi = 26_{-10}^{+11}$$^\circ$ measured by \citetalias{WeissermanGillis2025}. As such, we do not find strong evidence for a substantial misalignment of the stellar obliquity of TOI-3714.

\subsection{Constraints on Additional Planetary, Substellar, and Stellar Companions}\label{sec:companions}

Previously, \citetalias{WeissermanGillis2025} suggested that KL migration driven by a distant stellar companion is expected to be highly efficient around early M dwarfs compared to more massive stars. However, while any distant massive companion, be it stellar, substellar, or planetary mass, could potentially drive KL migration, only two out of the five short-period GEMS with RM measurements in the literature are in wide binary systems capable of driving HEM (i.e. TOI-5293 A b, \citealt{CanasKanodia2023}; TOI-3714 b, \citealt{CanasKanodia2022}). Here we calculate the detection sensitivity to distant massive companions in existing observations of all five known GEMS with RM effect measurements, including characterizing the sensitivity of existing RV, high-contrast speckle imaging, and astrometric data of TOI-5205, TIC 46432937, TOI-4201, TOI-5293, and TOI-3714.

\subsubsection{RV Sensitivity}\label{subsec:companions-rv}

To characterize the sensitivity of archival multi-epoch RVs, we perform a suite of injection-recovery tests on the RV data presented in each short-period GEMS' discovery paper. \citet{KanodiaMahadevan2023} obtained seven RVs of TOI-5205 between April 20, 2022 and May 19, 2022 (a baseline of 29 days) using the HPF spectrograph with a median RV uncertainty of $20.3$ m/s. \citet{HartmanBayliss2024} obtained eight RVs of TIC 46432937 between November 24, 2022 and February 22, 2023 (a baseline of 94 days) using ESPRESSO, with a median RV uncertainty of $2.0$ m/s. \citet{HartmanBakos2023} obtained twelve RVs of TOI-4201 between September 7, 2022 and January 10, 2023 (a baseline of 125 days) using the Keck/HIRES spectrograph, with a median RV uncertainty of $7.6$ m/s. \citet{CanasKanodia2023} obtained sixteen RVs of TOI-5293 between September 10, 2022 and December 1, 2022 (a baseline of 61 days) using the HPF spectrograph, with a median RV uncertainty of $29.4$ m/s. \citet{CanasKanodia2022} obtained 12 RVs of TOI-3714 from HPF between August 24, 2021 and December 23, 2021 with a median RV uncertainty of $23$ m/s and 8 RVs from the NEID spectrograph between September 22, 2021 and January 8, 2022 with a median RV uncertainty of $11.5$ m/s (with a total baseline across both spectrographs of 137 days).

We subtract the best-fit planetary and instrumental models from each RV time series using the parameters presented in each paper. This includes linear trends, offsets, and the addition of white noise jitter when applicable. We then inject $2 \times 10^6$ Keplerian signals of hypothetical external companions across a wide range of parameter space. These signals are described by {six} parameters: the companion mass $M_\mathrm{comp}$, uniformly drawn in log space on $[0.01, 1000] M_J$ (i.e. roughly 3 Earth masses to a solar mass); the companion semimajor axis $a$, uniformly drawn in log space on $[0.1, 10]$ AU; the epoch $t_0$, uniformly drawn from the full phase of the injected orbit; {the companion inclination $i$, where planetary-mass $(M_\mathrm{comp} < 13 M_J)$ companions are assumed to be coplanar for simplicity and more massive companions have isotropic inclinations; the argument of periapse $\omega$, drawn uniformly on $[0, 2\pi)$; and the companion eccentricity $e$,  drawn from different beta distributions depending on the mass regime ($\beta(1.06,3.68)$ when $M_\mathrm{comp} < 13 M_J$, \citealp{RosenthalHoward2024}; $\beta(1.879,3.2.470)$ when $13 M_J < M_\mathrm{comp} < 75 M_J$, \citealp{FerreiraRice2026}; and $\beta(0.704, 0.573)$ when $M_\mathrm{comp} > 75 M_J$, \citealp{FerreiraRice2026})}.

For each injection, we quantify a successful recovery by calculating the difference in Bayesian information criterion ($\Delta$BIC) that compares a linear model against a Keplerian model that encodes the companion-induced signal. For the latter case, we assume that the best-fit values of the {six} Keplerian parameters are consistent with the injected values (meaning a total of {six} model parameters: a flat offset, $M_\mathrm{comp} \sin i$, $a$, $t_0$, {$\omega$, and $e$}). For the former model, we perform least squares minimization to find the linear relation, described by a slope and a constant offset, that fits the injected RVs the best. We then classify an injected signal as being detected if $\Delta {\rm BIC} \leq -20$. We then define the contour in mass-separation space that separates detections from nondetections by splitting the injections up into 400 equal bins in log-semimajor-axis space and fitting the masses and semimajor axes to a logistic curve, taking the value of the mass at a given point on the sensitivity contour to be the {point on the} logistic curve {corresponding to a detection probability of $90\%$}. Our RV sensitivity curves for each planet are depicted in Figure \ref{fig:sensitivity-curves}.

\subsubsection{Speckle Imaging Sensitivity}\label{subsec:companions-speckle}
TOI-5205, TOI-4201, TOI-5293, and TOI-3714 have archival speckle imaging datasets that we use to complement the RV datasets in Section~\ref{subsec:companions-rv} in constraining the presence of widely-separated massive companions in these systems. These datasets include $z'$-band speckle imaging of TOI-5205, TOI-5293, and TOI-3714 with the NN-EXPLORE Exoplanet Stellar Speckle Imager (NESSI) at the WIYN telescope \citep{KanodiaMahadevan2023,CanasKanodia2023,CanasKanodia2022} and speckle imaging of TOI-4201 with Gemini-South/Zorro in the 562 nm and 832 nn bands \citep{HartmanBakos2023}. The contrast curves presented therein describe the sensitivity to nearby companions via high-contrast imaging.

We begin by converting the contrast curves from their native units of companion $\Delta m$ versus angular separation to companion mass versus projected separation in au. We convert from angular to projected separations using each star's known distance from Gaia DR3 \citep{BailerJones2021}. We then convert each imaging passband's $\Delta m$ between the host star and hypothetical companions to companion masses based on synthetic photometry from the AMES-COND 2000 stellar/substellar/giant exoplantary atmosphere models \citep{Allard2001}. We do so by calculating synthetic photometry using the Spanish Virtual Observatory's Synthetic Photometry service\footnote{\url{https://svo2.cab.inta-csic.es/theory//newov2/syph.php}} over $T_{\rm eff} \in [300,4000]$ K in 100 K intervals and $\log{g} \in [3.5,5.5]$ in intervals of 0.5 dex. We then map each $\{ T_{\rm eff}, \log{g} \}$ combination to an object mass using the AMES-COND 2000 isochrones compiled for the \texttt{species} Python package \citep{Stolker2020}. For TOI-5205, we assume a fixed field age of 5 Gyrs given the lack of existing age measurements for the system, while for TOI-4201, we assume the reported age of 1.4 Gyrs \citep{GanCadieux2023}. {As TOI-5293 and TOI-3714 have published age ranges of 0.7-5.1 Gyrs \citep{CanasKanodia2023,CanasKanodia2022}, we assume ages of 2.9 Gyrs corresponding to the center of these age ranges.} To ensure a 1-on-1 comparison between the planet host stars and hypothetical companions, we repeat these steps for the closest matching AMES-COND 2000 atmosphere model of each host star. We then use the synthetic photometry to construct a mapping between companion mass and $\Delta m$. We interpolate this function at the $\Delta m$ points in each contrast curve to derive contrast curves in units of companion mass versus projected separation. Our speckle imaging contrast curves are included in Figure~\ref{fig:sensitivity-curves}.

\subsubsection{Astrometric Sensitivity}\label{subsec:companions-astrometry}
To search for companions at intermediate separations between our RV and speckle imaging constraints, we fit existing astrometric data from Gaia Data Release 2 (DR2) and 3 (DR3) using the same methodology as in the calibrated joint Gaia-Hipparcos astrometric catalog \citep[G23H;][]{Thompson2026}. However, we note that none of the five GEMS systems that we analyze have Hipparcos data. G23H reports calibrated DR2 and DR3 proper motion anomalies in the DR3 reference frame, in combination with the Hipparcos-Gaia Catalog of Accelerations (HGCA), for the proper motion and scaled position differences calibrated against the DR3 reference frame \citep{hgca_dr3}. The G23H framework also relies on the DR3 astrometric excess noise derived from independently calculated noise estimates following the method of \citet{Kiefer2025} and Gaia radial velocity uncertainties from the \texttt{paired} catalog \citep{paired}. These two noise parameters correspond to unaccounted-for signals in Gaia's single-object astrometric solutions, indicating the possible presence of unseen companions. In particular, the \texttt{paired} data are used to rule out close stellar binaries by a lack of significant RV variability. We fit the calibrated astrometric measurements and excess noise parameters using the comprehensive orbit modeling package \texttt{Octofitter} \citep[v8.2.5;][]{Thompson2023} to place detection limits on unseen companions. 

We use an MCMC to marginalize over the DR2 and DR3 sampling epochs and calibration uncertainties, providing the tightest constrains to date on massive companions in the absence of epoch astrometry that will ultimately be released in Gaia Data Release 4. We note that none of our targets were observed by Hipparcos, such that we fit the calibrated Gaia DR2-DR3 proper motion anomaly and excess noise uncertainties from G23H implemented in \texttt{Octofitter}. We fit the orbital eccentricity $e$, argument of periastron $\omega$, orbital inclination $i$, the longitude of ascending node $\Omega$, and the fraction of orbit past periastron $\tau$. with wide uniform priors, and fit the orbital period $\log{P}$ (log-uniform), mass ratio $q=M_{\rm comp}/M_\star$ (log-uniform) with wide log-uniform priors. TOI-3714, TIC 46432937, and TOI-5293 have known spatially resolved co-moving companions, which are already accounted for by \texttt{Octofitter} when it models any excess astrometric and radial velocity noise. 
The astrometric sensitivity curves are included in Figure \ref{fig:sensitivity-curves}. 




\subsubsection{Sensitivity Results}\label{subsec:companions-results}

\begin{figure}[!h]
    \centering
    \includegraphics[width=.97\linewidth]{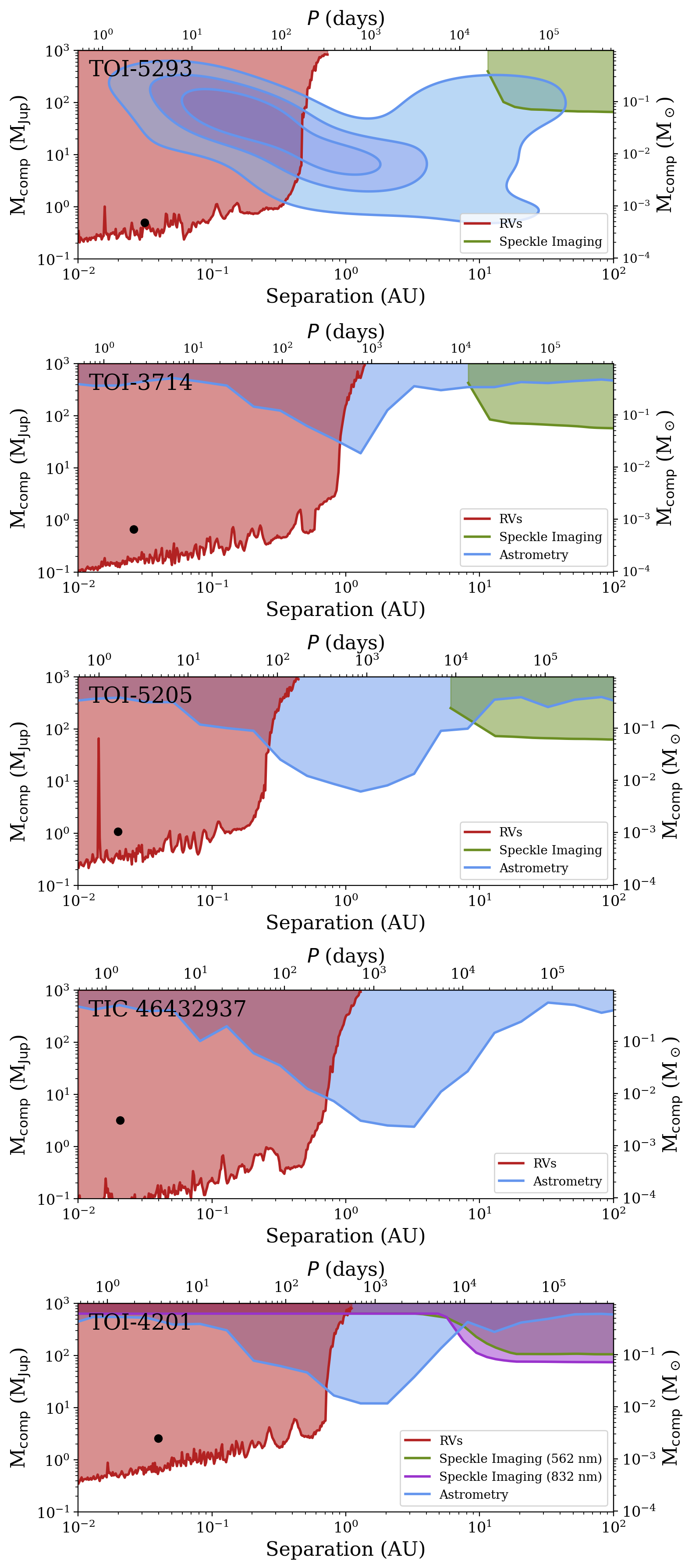}
    \caption{Detection sensitivity curves to massive companions as a function of companion mass and projected separation in the planetary systems around all five GEMS hosts with existing RM measurements. In TOI-5293's case, the blue contours represent the sampled parameter space corresponding to a possible second companion. The sensitivity curves are derived from archival RVs (red), speckle imaging (green), and Gaia DR2+DR3 astrometric data (blue). The black dots denote the masses and radii of the known HJs in each system.}
        \label{fig:sensitivity-curves}
\end{figure}

Our full suite of sensitivity curves {from RVs, astrometry, and speckle imaging} are compiled in Figure~\ref{fig:sensitivity-curves}. We find no evidence for additional massive companions around TOI-5205, TIC 46432937, TOI-4201, and TOI-3714. However, we highlight that these data are insufficient to rule out companions in large regions of the $a-M_{\rm comp}$ parameter space. While the RVs can rule out companions of Jupiter-mass or greater out to a $\sim 0.1$ au for all systems, they cannot rule out planetary-mass companions (i.e. $\lesssim 13\, M_{\rm Jup}$) beyond $\sim 0.3- 1$ au. While astrometry and speckle imaging rule out stellar-mass companions beyond $\sim 10$ au for all four systems with speckle imaging, these systems could still harbor objects in the brown dwarf regime ($\approx 10-100\, M_{\rm Jup}$) beyond several au. Much of each system's $a-M_{\rm comp}$ parameter space therefore remains unexplored by archival data, particularly substellar or Jovian-mass companions that could still be capable of driving HEM via KL interactions or gravitational scattering events early on in the system's history. As such, we cannot rule out the possibility that TOI-5205 b, TIC 46432937 b, nor TOI-4201 b {underwent} HEM, despite not having any known massive companions.

The astrometric fit of TOI-5293 shows tentative evidence for another, previously unknown, companion in the system. While the astrometric posterior is suggestive of the presence of a substellar-mass companion ($\sim 10-100\, M_{\rm Jup}$) between $\sim 0.1-2$ AU, the solution is poorly constrained with tails out to both smaller and large separations. We therefore cannot confirm nor characterize this putative companion in detail.
Fortunately, much of the posterior parameter space constrained by our astrometric solution is ruled out by the RV and speckle imaging data (Figure \ref{fig:sensitivity-curves}). However, a companion with a mass between $\sim1-20\, M_{\rm Jup}$ at or beyond $\sim$ 0.5 au remains consistent with all available RV, speckle imaging, and astrometric data. Forthcoming epoch astrometry data from Gaia DR4 is anticipated to improve astrometric sensitivity and clarify our interpretation of the {tentative} companion around TOI-5293 A.



\subsection{{Feasibility of Kozai-Lidov Migration for GEMS}}\label{subsubsec:companions-kl}

Planets orbiting stars with a wide massive companion will be acted on by the KL mechanism if the initial inclination between the orbital planes of the planet and companion is sufficiently large \citep[i.e. $>39.2^\circ$;][]{InnanenZheng1997}. This angular momentum exchange process produces large periodic oscillations between the planet's eccentricity and inclination. This effect, when combined with strong tidal damping near periapse, shrinks the planet's orbit and could be responsible for producing close-in GEMS with low obliquities.

\begin{figure*}[!ht]
    \centering
    \includegraphics[width=\linewidth]{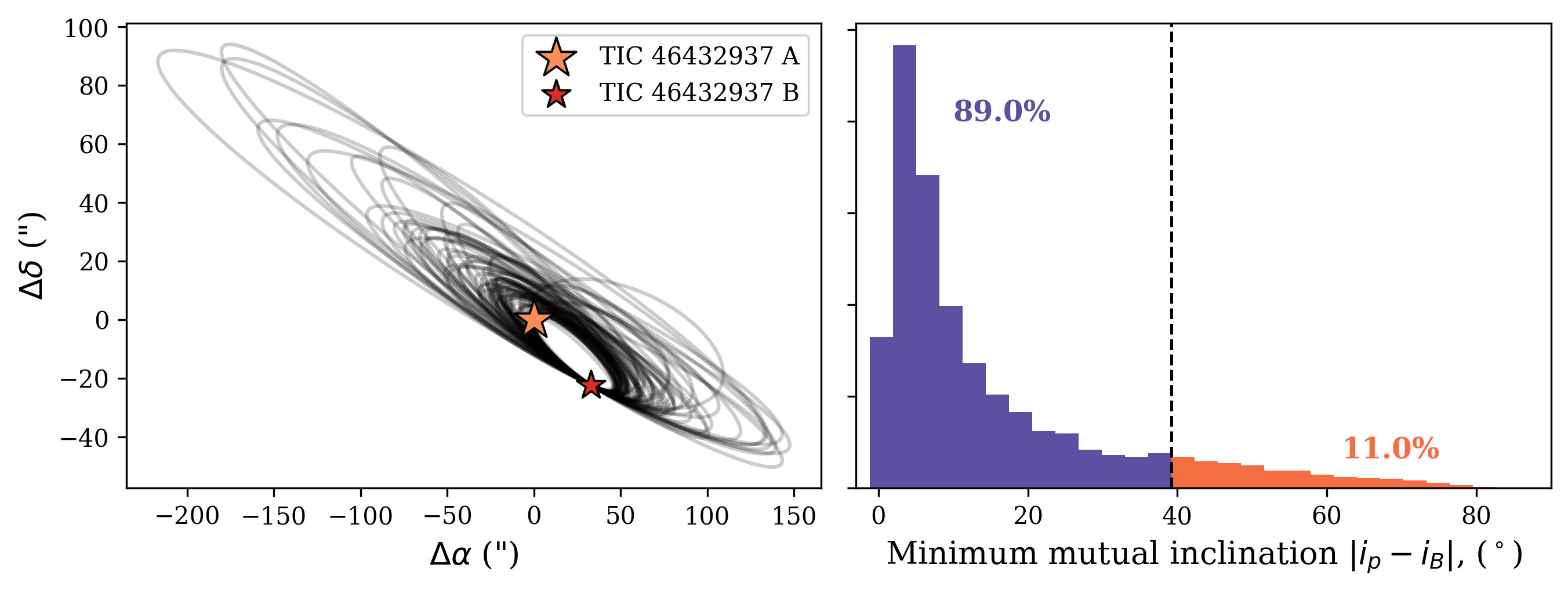}
    \caption{Results of orbital fits to Gaia DR3 astrometric data for the TIC 46432937 system. The left column shows a random selection of sky-projected binary orbit solutions from our astrometric fit posteriors, while the right column shows the distribution of minimum mutual inclinations between each system’s planet and binary companion. The vertical dashed line at $\arccos \sqrt{3/5} \approx 39.2^\circ$ corresponds to the minimum mutual inclination needed to drive KL migration. The percentiles of the mutual inclination distribution above and below $39.2^\circ$ are annotated and depicted by different colors.}
    \label{fig:mutual-inc}
\end{figure*}

Recently, \citet{GanLHeureux2026} identified a co-moving distant companion to TIC 46432937, Gaia DR3 2984391358868786560 (hereafter TIC 46432937 B for consistency with \citealt{GanLHeureux2026}). Here, we investigate whether this binary companion could possibly drive KL migration of TIC 46432937 b{, and check if the stellar companions of TOI-5293 and TOI-3714 could drive KL migration as well.}

{It has been proposed that the eccentric Kozai-Lidov (EKL) effect, a specific KL process in situations where the outer perturber is on an eccentric orbit \citep{Naoz2016}, could be responsible for HEM for these systems. To test whether the EKL effect could drive planet migration in these systems with stellar companions, we claim that EKL oscillations are feasible if the following three requirements are met:}

{\begin{itemize}
    \item The minimum mutual inclination of the planet and binary companion exceeds $39.2^\circ$.
    \item The quadrupole-level EKL precession timescale $t_{\rm quad}$ is shorter than the initial general relativistic apsidal-precession rate $\dot{\omega}_{\rm GR}$ such that EKL oscillations are not suppressed, i.e. $t_{\rm quad} \dot{\omega}_{\rm GR} \lesssim 1$ \citep{FabryckyTremaine2007}.
    \item The EKL descent time $t_{\rm desc}$, the time required for a planet on an initially wide orbit to reach a pericenter equal to the present-day pericenter of the HJ \citep{WeldonNaoz2024}, is less than the age of the system.
\end{itemize}}

{We check the first requirement by  constraining} the orbital inclination of the binary star system using Gaia DR3 astrometry. We use the \texttt{lofti\_gaia} package to fit the Keplerian orbital parameters of the binary system given input astrometry and the masses for both stellar components. Due to the absolute $K_S$ magnitude of this companion, we suspect that this companion is likely an M dwarf; as such, we estimate the mass of TIC 46432937 B using the empirically calibrated $M_\star - M_{K_S}$ relation for M dwarfs in \citet{MannDupuy2019} and recover a companion mass of $0.406 \pm 0.009\, M_\odot$. From our astrometric orbital fit we measure a companion orbital inclination of TIC 46432937 B $i_B$ of $76\pm 8 ^\circ$. We also find that the binary has a probable period of $P = 2.8_{-1.6}^{+1.3} \times 10^5{\rm  yr}$ and is eccentric ($e = 0.65_{-0.21}^{+0.33}$). A sample of posterior orbital solutions are shown in the left panel of Figure \ref{fig:mutual-inc}.

We then calculate the posterior of the minimum mutual inclination between the planet and the companion, $\Delta i = |i_p - i_B|$. Figure~\ref{fig:mutual-inc} depicts a selection of orbits in the sky plane based on random samples of the joint posterior of this fit, as well as a plot of the minimum mutual inclination distribution. We find a skewed minimum mutual inclination distribution, with a median value of $8.3$$^\circ$, and only $\sim 11\%$ of the $\Delta i$ posterior exceeds $39.2^\circ$. However, we note that $\Delta i$ is only the minimum mutual inclination between the planet and companion, as the position angle of the planet on the sky is unknown. Therefore, while the data are consistent with a low mutual inclination between the GEMS TIC 46432937 b and the wide stellar companion TIC 46432937 B, this does not on its own rule out that the system’s true mutual inclination may be sufficiently misaligned to drive KL migration. {TOI-3714 and TOI-5293 have already been checked in this way and shown to have sufficiently high mutual inclination for both targets \citep{WeissermanGillis2025}.}

{To test our second requirement, we calculate $t_{\rm quad}$ and $\dot{\omega}_{\rm GR}$ for fiducial initial orbital parameters of our giant planets. The characteristic quadrupole-level EKL timescale (i.e. the timescale of oscillations driven by the KL mechanism) is given by}

{\begin{equation}
    t_{\rm quad} = \frac{2 P_p^2}{3 \pi P_c} \frac{M_\star + M_p + M_c}{M_c} (1 - e_c^2)^{3/2},
\end{equation}}

\noindent {and the initial general relativistic apsidal-precession rate (i.e. the rate of pericenter precession due to relativistic effects) is given by}

{\begin{equation}
    \dot{\omega}_{\rm GR} = \frac{3 G^{3/2} (M_\star + M_p)^{3/2}}{a_p^{5/2} c^2 (1 - e_p)^2},
\end{equation}}

\noindent {where $M, P, a$, and $e$ denote the masses, periods, semimajor axes, and eccentricities of the bodies respectively, with subscripts $_\star,\, _p,$ and $_c$ denoting the primary star, the planet, and the perturbing companion, respectively. We assume conservative fiducial values of $a_p = 5\, {\rm  AU}$ and $e_p = 0$, as both will maximize the value of $\dot{\omega}_\textrm{GR}$, providing conservative estimates on whether EKL oscillation will be precessed. We randomly draw stellar and planetary parameters from our priors presented in Appendix \ref{ap:posteriors} and our companion parameters from the posteriors of our \texttt{lofti\_gaia} runs. We repeat this process $10^4$ times and observe the distributions of the resulting $t_{\rm quad} \dot{\omega}_{\rm GR}$ values. The distributions of these timescales, as well as the other timescales discussed in this section and Section \ref{subsec:disc-obldamp}, are included in Table \ref{tab:kl-timescales} below.}

\begin{table*}[]
    \centering
    {\begin{tabular}{c|c|c|c|c|c}
    \hline \hline System & $t_\textrm{quad}$ & $\dot{\omega}_\textrm{GR}^{-1}$ & $\tau_\lambda$ (Eq. \ref{eq:obldamp-1})
         & $\tau_\lambda$ (Eq. \ref{eq:obldamp-2}) & System Age\\ \hline
        TOI-5293 & $0.01_{-0.01}^{+0.07}$ & $0.92_{-0.03}^{+0.04}$ & $1740_{-700}^{+1580}$ & $16800_{-7270}^{+16300} $ & 0.7 - 5.1 \citep{CanasKanodia2023} \\
        TOI-3714 & $0.0001_{-0.0001}^{+0.0011}$ & $0.83 \pm 0.03 $ & $440_{-110}^{+100}$ & $3270_{-770}^{+850} $ & 0.7 - 5.1 \citep{CanasKanodia2022} \\
        TIC 46432937 & $1.1_{-1.0}^{+8.5}$ & $0.71_{-0.05}^{+0.06}$ & $10.7_{-0.5}^{+0.6}$ & $29 \pm 4$ & $7.4\pm 5.1$ \citep{HartmanBayliss2024} \\
        TOI-5205 & -- & -- & $360 \pm 50$ & $610_{-90}^{+130}$ & -- \\
        TOI-4201 & -- & -- & $360 \pm 30$ & $1070 \pm 130$ & $1.4_{-0.7}^{+0.6}$ \citep{GanCadieux2023} \\
    \end{tabular}
    \caption{The distributions of the reported quadrupole-level EKL timescale and (inverse) rate of pericenter precession due to relativistic effects for each system, as well as the obliquity damping timescales calculated in Equation \ref{eq:obldamp-1} and \ref{eq:obldamp-2} and system ages. All numbers are in Gyr.}}
    \label{tab:kl-timescales}
\end{table*}

{We find that $99\%$, $98\%$, and $43\%$ of the samples of TOI-3714, TOI-5293, and TIC 46432937, respectively, have $t_{\rm quad} \dot{\omega}_{\rm GR} < 1$. As such, while uncertainties regarding our stellar companion parameters prevent more concrete conclusions, it is likely that EKL oscillations are not suppressed by general relativistic effects in the TOI-3714 and TOI 5293 systems, while the same outcome is reasonably plausible for TIC 46432937. However, it should be noted that the uncertainty in these values are dominated by our uncertainties in $P_c$ and $e_c$, which can cause $t_{\rm quad} \dot{\omega}_{\rm GR}$ to vary by up to two orders of magnitude for all three systems, so more concrete conclusions cannot be easily drawn.}

{To test our third requirement, we draw $10^4$ initial configurations of stellar, planetary, and companion parameters, and calculate $t_{\rm desc}$ for each. We follow the procedure to calculate $t_{\rm desc}$ described in Appendix D of \citet{ShariatElBadry2026}. This includes retaining only sampled configurations that are dynamically stable, overcome initial GR precession, and fall within the calibration range of the analytic $t_{\rm desc}$ expression. The calculation relies on host star and planetary masses that we sample from our priors in Appendix \ref{ap:posteriors}, while companion masses and orbital parameters are drawn from our \texttt{lofti\_gaia} sample outputs. We assumed that the planet's original semi-major axis was drawn from a log-uniform distribution between $[1, 10]$ AU and its eccentricity was sampled from the $\beta(0.844,3.313)$ distribution \citep{FerreiraRice2026}.}

{We find $t_{\rm desc}$ timescales for TOI-3714 b, TOI-5293 b, and TIC 46432937 b of $0.009_{-0.008}^{+0.056}{\rm  Gyr}$, $0.3_{-0.2}^{+0.9}{\rm  Gyr}$, and $9.1_{-6.1}^{+12.3}{\rm Gyr}$, respectively. $99\%$, $95\%$, and $42\%$ of EKL-favorable configurations have descent timescales below these system's median ages. Because the descent timescales of the majority of the configurations for TOI-3714 and TOI-5293 are much less than the ages of these systems, we conclude that the known stellar companions to TOI-3714 and TOI-5293 are capable of inducing EKL migration of their respective GEMS. TIC 46432937 b's descent timescale is long, largely due to its companion's large orbital separation, but does have large uncertainties driven by uncertainties in the companion's orbital parameters. In particular, if the planet forms further out than our $5 \textrm{ AU}$ fiducial estimate, or if TIC 46432937 B is on a highly eccentric orbit, the descent timescale can be on the order of hundreds of Myr. While it is far less likely that TIC 46432937 B can induce high-eccentricity migration for TIC 46432937 b through EKL oscillations than the other two systems, we cannot fully rule out this possibility with our current, loose constraints on the binary companion from Gaia DR3.}




\subsection{{Feasibility of Obliquity Damping for GEMS}}\label{subsec:disc-obldamp}

{In addition to testing if EKL migration is feasible, we test whether obliquity damping by stellar tides could feasibly align all the five GEMS with RM detections.} For each planet, we retrieve stellar and planetary parameters as before and calculate the obliquity damping timescale from \citet{EggletonKiseleva-Eggleton2001} and \citet{WinnFabrycky2010}:

\begin{equation}\label{eq:obldamp-1}
    \tau_\lambda = k \left(\frac{M_{cz}}{M_p}\right) \left(\frac{a}{R_\star}\right)^6 \left(\frac{(1 - e^2)^{9/2}}{1 + 3e^2 + \frac{3}{8}e^4} \right),
\end{equation}

\noindent where $k$ is a constant, $M_{cz}$ is the mass of the stellar convective zone, $M_p$ is the planet mass, $a$ is the planet’s semi-major axis, $R_\star$ is the stellar radius, and $e$ is the planet’s orbital eccentricity. We follow \citet{RiceWang2022} by fixing $k = 10^3$ {years} in our calculations. {It should be noted that \citetalias{WeissermanGillis2025} uses an incorrect constant of $k = 10^3$ seconds in their analogous calculations of $\tau_{\lambda}$ for TOI-3714 and TOI-5293, and find incorrect timescales as a result; we recalculate the obliquity damping for these planets herein.} We assume $e = 0$ for all GEMS where eccentricity is not reported (as this maximizes estimates for $\tau_\lambda$). In order to calculate $M_{cz}$, we adopt the convective zone mass calculations for low-mass stars from \citet{AmardPalacios2019}, interpolating over the reported convective zone masses at {the reported age of each system (or, where not applicable, a field age of 5 Gyr)} across relevant stellar masses and metallicities.

{Our calculated timescales are reported in Table \ref{tab:kl-timescales}. We calculate long obliquity damping timescales for all five GEMS with RM effect measurements, ranging from $10.7-1740$ Gyrs. To confirm these values, we also calculate the obliquity damping timescales using a different prescription \citep{Zahn1977,RiceWangWang2022}:}

{\begin{equation}\label{eq:obldamp-2}
    \tau \approx 10\, {\rm  Gyr} \left(\frac{a/R_\star}{40}\right)^6 \left(\frac{M_p}{M_\star}\right)^{-2},
\end{equation}}

\noindent {Our calculated timescales are reported in Table \ref{tab:kl-timescales}. In this prescription, we find similarly long damping timescales, ranging between $29-16,800$ Gyr and consistent to within an order of magnitude with the prior damping timescales. We note that this scaling relation is calculated using stellar binaries in mind, assuming planets realign via a similar process. Using both prescriptions of tidal damping in Eqs.~\ref{eq:obldamp-1} and \ref{eq:obldamp-2}, all five GEMS known to have aligned obliquities are in systems whose tidal damping timescales well exceed the system's age by a factor of $\sim 1-6000\times$. This may suggest that all known aligned GEMS could not be tidally realigned if they entered their current orbits with high orbital obliquities.}

{An alternative measure of the feasibility of tidal damping for these GEMS compares the rotational angular momentum of the stellar convective layer $L_{\rm conv}$ and the planet's orbital angular momentum $L_p$, to assess whether these GEMS have sufficient angular momentum to realign their host stars. We calculate these two values as}

{\begin{equation}
    L_{{\rm conv}} = \frac{2\pi}{P_{\rm rot}} \kappa M_{\rm cz} R_\star^2
\end{equation}}

\noindent {and}

{\begin{equation}
    L_p = \sqrt{GM_\star M_p^2 a (1 - e)^2} \cos i_p,
\end{equation}}

\noindent {where $P_\textrm{rot}$ is the stellar rotation period and $\kappa = 0.1$ is the fixed moment of inertia coefficient \citep{BaraffeHomeier2015}. We assume that $e = 0$, and for stars without known rotation periods, we assume a fiducial value of $P_{\rm rot} = 30\textrm{ days}$. We find angular momentum ratios $L_{\rm conv}/L_p$ of less than unity for all GEMS, ranging from $0.02-0.76$. As such, while the estimated timescales to realign the obliquities of all five GEMS systems appear prohibitively long using the aforementioned tidal realignment prescriptions, the planetary orbits feasibly have sufficient angular momentum to realign the spin axes of their host stars.}

\subsection{Implications for Short-Period GEMS Formation}\label{subsec:disc-implications}


{One interpretation of the prohibitively long damping timescales reported in Table~\ref{tab:kl-timescales} is that the low measured obliquities of GEMS are primordial obliquities. If this is the case, then the ubiquitously aligned obliquities of close-in GEMS can be interpreted as evidence for disk migration over HEM, as HEM would not be expected to produce aligned planets in the absence of efficient tidal damping. However, this interpretation would stand at odds with circumstantial evidence that HEM is the dominant formation mechanism for GEMS, including that wide binary companions to M dwarfs are more common in systems with GEMS \citep{GanLHeureux2026,MessamahBouchy2026} at a rate consistent with HEM plus efficient damping \citepalias{WeissermanGillis2025}, and that we see a mass-ratio dependence on obliquities for planets around M dwarfs, where high-mass ratio systems such as GEMS tend to be aligned \citep{WinnFabrycky2010,RiceWang2022,ZanazziDewberry2024,GanWang2024,RusznakWang2025} while lower-mass planets around M dwarfshave a wider range of obliquities \citep[e.g.][]{BourrierZapateroOsorio2022,Libby-RobertsSchutte2023,Carteret2026}. \emph{If} disk migration is the dominant cause of GEMS migration, then one must explain these trends in the absence of HEM. Perhaps the presence of stellar companions may enhance gas giant formation through inducing high-density spiral arms that could increase planet formation efficiency \citep{WagnerApai2015,ReynoldsTobin2021}, but whether these stellar companions enhance giant planet formation around M dwarfs remains an open question (Gillis et al. in prep).} 

{On the other hand, if efficient tidal damping is indeed responsible for these planets' obliquities, then one must explain how obliquity damping happened within the lifetime of the systems. One possibility is that Eq.~\ref{eq:obldamp-1} is an insufficient prescription for calculating tidal damping timescales. Indeed, simulations of tidal realignment being limited to convective zones around Sun-like stars (so-called skin deep or decoupled tidal realignment) suggest that nearly all HJs around these stars should be aligned \citep[see Fig. 23 of ][]{AlbrechtDawson2022} despite the fact that using the prescription in Eq.~\ref{eq:obldamp-1} should result in $\tau > 10^{11}\textrm{ yr}$ for many HJs around Sun-like stars (see Fig. 3 of \citet{RiceWang2022}). This implies that Eq.~\ref{eq:obldamp-1} is providing an incomplete picture of the rate at which obliquities are damped by tides. Alternatively, other models of tidal realignment may be responsible for obliquity damping, such as tidal dissipation of inertial waves \citep{Lai2012,LinOgilvie2017} or tidal dissipation being steeply dependent on the forcing frequency \citep{PenevBouma2018,AndersonWinn2021}, that may be able to damp obliquities at a much faster rate. However, as these models have focused primarily on their predictions of tidal damping around Sun-like stars, further work is necessary to understand if these models can explain tidal damping of GEMS.}

\subsection{Future Observations}\label{subsec:disc-future-observation}

With a sample of just five close-in GEMS with measured stellar obliquities that are all well-aligned, their dominant formation and migration channel remains an open question. If the low obliquities of known close-in GEMS are indeed due to M dwarfs efficiently damping any misaligned obliquities, then it is possible all close-in GEMS are expected to have low obliquities at old ages, regardless of whether these planets migrated via HEM or from disk-driven migration. If tidal damping is ubiquitous, then finding more aligned GEMS will not necessarily provide any new information to distinguish between primordial alignment and tidal realignment. As such, alternative observing strategies will be needed to robustly distinguish between GEMS formation mechanisms, either on an individual planet basis or on a population level. Here we propose future observing campaigns that seek to establish the dominant GEMS migration pathway.

\subsubsection{RM Observations of GEMS around Late M Dwarfs}\label{subsubsec:future-observation-late-m}

While all GEMS with RM effect observations appear to be aligned, all such planets orbit relatively massive M dwarfs ($0.39-0.61\,M_\odot$). Recent theoretical studies of obliquity damping modeling resonance locks with non-axisymmetric modes, predict that if HJs orbiting mid-to-late M dwarfs do form via HEM, resonance locking should cause obliquities to rise back up around stars with $T_{\rm eff}\lesssim 3400$ K \citep{ZanazziChiang2025}. Measuring the obliquities of GEMS around cool M dwarfs \citep[e.g. TOI-6894 b, with $T_{\rm eff} = 3007{\textrm{ K}}$;][]{BryantJordan2025} should serve as a useful test of this prediction.

\subsubsection{Searching for Inner Companions to {Close-in} GEMS}\label{subsubsec:future-observation-friends}



{HEM is dynamically disruptive in multi-planet systems. It is therefore expected that inner companions to HJs formed via HEM will not survive the process, such that HJs should preferentially exist in low-multiplicity systems \citep{MustillDavies2015}. While a lack of confirmed inner companions does not on its own prove that HEM is the dominant HJ migration process \citep{RadzomWang2026}, the discovery of inner companions to {close-in} GEMS would provide strong counter evidence for HEM as the dominant migration mechanism. Currently, no {close-in} GEMS system has a confirmed inner planetary companion such that HEM plausibly remains the dominant migration channel for {close-in} GEMS. Deep transit and radial velocity searches for inner planets in {close-in} GEMS systems would serve to test predictions of HEM being the predominant migration mechanism for GEMS.}

\subsubsection{Atmospheric Characterization}\label{subsubsec:future-observation-atmospheric}

An alternative pathway to uncover the origins of GEMS would be to probe their atmospheric composition and link these to predictions from planet formation models \citep[e.g.][]{BitschSchneider2022,TurriniSchisano2021,PenzlinBooth2024}. Atmospheric diagnostics of planetary formation location have long focused on the C/O ratio, but it now known to lead to degenerate results when not considered alongside the planet's accretion and evolutionary history \citep[e.g.][]{TurriniSchisano2021,MolliereMolyarova2022,BitschSchneider2022}. However, the combination of elemental ratios (e.g. both C/H and O/H), refractory/volatile ratios (e.g. O/Si), and bulk atmospheric metallicity measurements allows the variety of formation pathways to be tested \citep[e.g.][]{HeBitsch2026}. Disk evolution models and observations indicate that inner disk regions around M dwarfs should be preferentially carbon-rich and water-poor \citep{PascucciHerczeg2013,MahBitsch2023}, indicating that planets forming in these regions should accrete gas with super-stellar C/O ratios. Uncovering population-level atmospheric compositions of GEMS with facilities like JWST and Ariel \citep[][O. Venot et al. in press]{Changeat_2025} could thus help to understand the diversity of their origins.

\subsubsection{Short-Period GEMS RM Effect Observations}\label{subsubsec:future-observation-short-period-gems}

While the discovery of more GEMS with aligned stellar obliquities would not distinguish between potential formation and evolutionary pathways, future detections of the RM effect for short-period GEMS will enhance the comparison of the obliquity distribution of close-in GEMS to that of HJs around hotter stars.

\begin{figure*}[ht]
    \centering
    \includegraphics[width=0.9\linewidth]{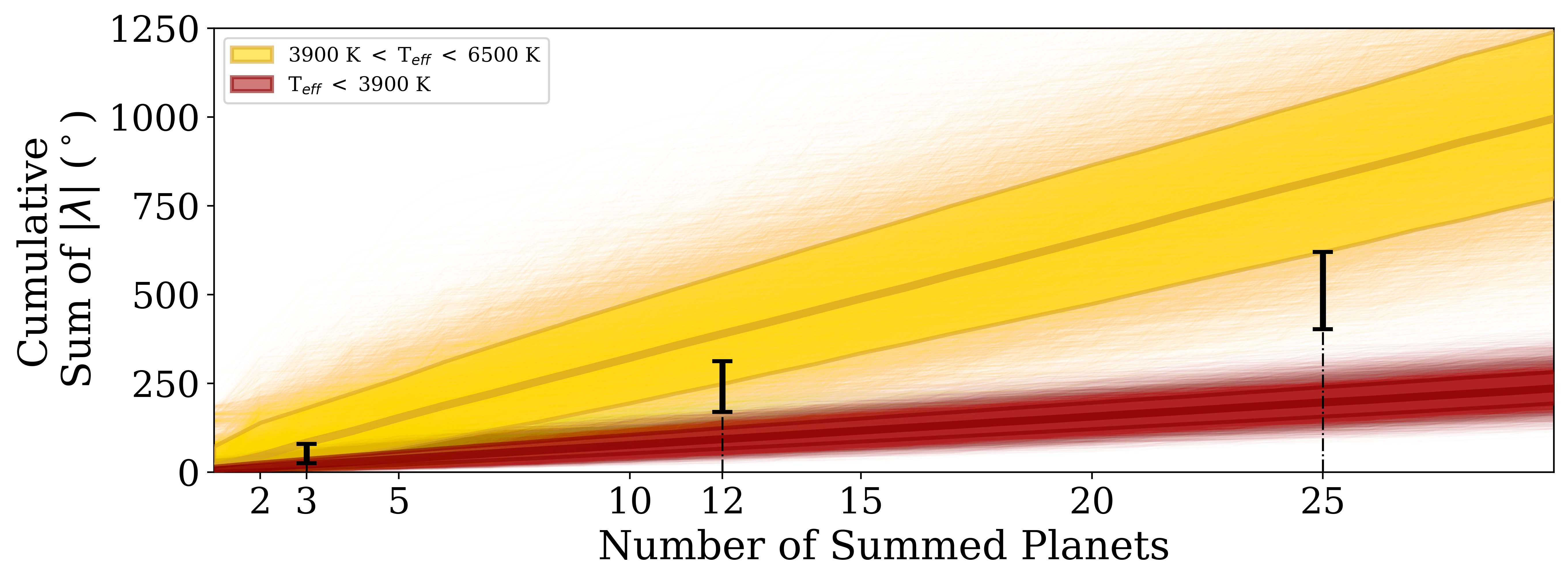}
    \caption{Comparison of the cumulative sky-projected stellar obliquities $|\lambda|$ for HJs around M dwarfs ($T_\mathrm{eff} < 3900\mathrm{ K}$; red) versus around stars below the Kraft break ($3900\mathrm{ K} < T_\mathrm{eff} < 6100\mathrm{ K}$; gold) as a function of number of planets with dections of the RM effect. The median and $1\sigma$ uncertainties on cumulative $|\lambda|$ are plotted and shaded, respectively. The combined uncertainties are plotted as black errorbars at the numbers of planets needed to distinguish between the two distributions at  1, 2, and 3$\sigma$. Individual samples from obliquity draws are overplotted with high transparency.}
    \label{fig:cumulative-lambdas}
\end{figure*}

Here we estimate the number of short-period GEMS needed to robustly distinguish between the sky-projected obliquity distributions $\lambda$ of short-period GEMS versus HJ-hosting AFGK stars. We compare the cumulative sums of absolute spin-orbit angles $|\lambda|$ from each stellar sample. {We retrieve the obliquities and associated uncertainties of all HJs whose host stars have effective temperatures below $6500\textrm{ K}$ from the Exoplanet Archive.} For increasing $N \in [1,30]$, we draw $|\lambda|$ for $N$ randomly-selected HJs from each stellar sample. Because the sample of short-period GEMS with measured $|\lambda|$ is five, we sample $N>5$ by assuming that the current $|\lambda|$ distribution from those five GEMS is representative of the underlying population. We compute the cumulative sum of $|\lambda|$ as a function of $N$. We repeat this process for $10^4$ realizations and measure the mean and standard deviations of the cumulative $|\lambda|$ estimates from all realizations as a function of $N$. The results are shown in Figure \ref{fig:cumulative-lambdas}.

We find that robustly distinguishing the $|\lambda|$ distributions around M dwarfs versus AFGK stars at the $2\sigma$ level will require a total of {12} short-period GEMS with RM detections, assuming that the recovered values of $|\lambda|$ are consistent with the current distribution of aligned GEMS. Similarly, increasing the significance to $3\sigma$ would require {25} total RM measurements. Establishing the degree of consistency in the obliquity distributions of short-period GEMS with HJs around earlier type stars can be compared to predictions of HEM and tidal damping theory, and so a comparison of statistically significant samples of RM measurements of GEMS has the potential to test whether GEMS share a common origin with other HJs.

\section{Summary \& Conclusion}\label{sec:conc}

We have presented detections of the RM effect using the MAROON-X spectrograph for three close-in giant planets around the early M dwarfs: TOI-3714, TOI-5205, and TIC 46432937. Our measurements of TOI-5205 and TIC 46432937 represent the fourth and fifth measurements of the RM effect for GEMS in the literature. We summarize our main conclusions below.

\begin{itemize}
    \item We detect the RM effect for all three planets at {$8\sigma-16\sigma$}. We measure sky-projected stellar obliquities of {$\lambda = 0_{-5}^{+6}$$^\circ$, $0\pm 3^\circ$, and $12_{-7}^{+9}$$^\circ$}, and projected stellar rotation velocities of {$v\sin {i_\star} = 0.53 \pm 0.04$ km/s, $1.32_{-0.18}^{+0.20}$ km/s, and $1.09_{-0.09}^{+0.10}$ km/s}, for TOI-5205, TIC 46432937 b, and TOI-3714, respectively.
    \item We compute a deprojected stellar obliquity of {$\psi = 29\pm 10^\circ$ for TOI-3714}, consistent with being aligned and consistent with previous measurements from \citetalias{WeissermanGillis2025}. While TOI-5205 and TIC 46432937 both have candidate or reported rotation periods in the literature, we believe neither of them are true rotation periods, and so deprojected stellar obliquities cannot be calculated for these two stars.
    \item The TOI-5205, TIC 46432937, and TOI-4201 systems do not show evidence for massive companions capable of driving HEM. However, existing RV, imaging, and astrometry data cannot rule out large portions of parameter space, and the presence of distant Jovian or sub-stellar companions beyond $\sim 1{\rm  au}$ remains plausible in all five GEMS systems with RM detections.
    \item Astrometric data for TOI-5293 shows tentative evidence for an additional companion between $\sim 10-100\, M_{\rm Jup}$ at $\sim 1-10$ AU. However, more data are needed to improve the interpretation of this result.
    \item Our findings present further evidence that short-period GEMS, both with and without distant companions, are preferentially aligned.
    {\item We find that these GEMS likely have sufficient angular momentum to realign their host stars, though the commonly used prescriptions of tidal damping suggest that it would take prohibitively long to realign these stars.}
    {\item While circumstantial evidence supports HEM as the primary driver of GEMS migration, these results suggest that how these planets were damped during the system's lifespan remains an open question.}
\end{itemize}


\begin{acknowledgments}
We thank Erik Gillis and Rohan Kumar for their scientific discussions regarding the rotation period of TOI-5205, Teo Moncik for their scientific discussions regarding the observations and discussions presented in this work, William Thompson for consultations regarding the interpretation of our astrometric fits{, Cheyenne Shariat for their discussions regarding KL migration feasibility, and Malena Rice for their discussions regarding tidal damping}.

DW acknowledges the support of the Natural Sciences and Engineering Research Council of Canada (NSERC).

RC acknowledges the support of the Natural Sciences and Engineering Research Council of Canada (NSERC) and the Canada Research Chairs (CRC) Program [Award ID: CRC-2024-00372]. 

The MAROON-X group acknowledges funding from the David and Lucile Packard Foundation, the Heising-Simons Foundation, the Gordon and Betty Moore Foundation, the Gemini Observatory, and the NSF (award number 2108465).

\end{acknowledgments}

\clearpage

\appendix

\section{RV Data}\label{ap:rv-data}

Table \ref{tab:ap-rvs} details the RVs for all {six} transits modeled in this work, including the one transit of TOI-3714 originally presented in \citetalias{WeissermanGillis2025}.

\begin{table}[h]
    \centering
    \caption{Combined RV measurements for all RM effect measurements fit in this work.}
    \begin{tabular}{cccc|cccc}
        \hline\hline Star & BJD & RV (m/s) & $\sigma$ (m/s) & Star & BJD & RV (m/s) & $\sigma$ (m/s) \\ \hline
        TOI-5205 & 2460889.90878183 & 75.9 & 5.0 & TIC 46432937 & 2461036.95974708 & -48.4 & 3.3 \\
        TOI-5205 & 2460889.92037926 & 63.7 & 6.1 & TIC 46432937 & 2461036.97135567 & -74.3 & 4.0 \\
        TOI-5205 & 2460889.93197678 & 52.9 & 7.5 & TIC 46432937 & 2461036.98295275 & -108.1 & 4.8 \\
        TOI-5205 & 2460889.9435858 & 53.2 & 7.0 & TIC 46432937 & 2461036.99456133 & -159.2 & 5.3 \\
        TOI-5205 & 2460889.95518322 & 21.4 & 7.0 & TIC 46432937 & 2461037.00615831 & -203.5 & 3.1 \\
        TOI-5205 & 2460889.96678064 & -17.6 & 9.0 & TIC 46432937 & 2461037.01776688 & -242.2 & 2.7 \\
        TOI-5205 & 2460907.84732447 & 68.0 & 2.2 & TOI-3714 & 2460654.81720071 & 43.3 & 5.5 \\
        TOI-5205 & 2460907.85892164 & 54.0 & 2.1 & TOI-3714 & 2460654.82864392 & 31.3 & 5.4 \\
        TOI-5205 & 2460907.8705304 & 59.0 & 2.2 & TOI-3714 & 2460654.84027962 & 29.3 & 4.6 \\
        TOI-5205 & 2460907.88212756 & 44.2 & 2.6 & TOI-3714 & 2460654.85173333 & 25.0 & 4.5 \\
        TOI-5205 & 2460907.89372482 & -2.0 & 3.0 & TOI-3714 & 2460654.86339273 & 33.0 & 4.0 \\
        TOI-5205 & 2460907.90533357 & -26.1 & 2.9 & TOI-3714 & 2460654.87504123 & 38.2 & 3.9 \\
        TOI-5205 & 2460907.91693073 & -12.2 & 5.7 & TOI-3714 & 2460654.88675543 & 19.8 & 3.2 \\
        TOI-5205 & 2460907.92852788 & -31.8 & 3.2 & TOI-3714 & 2460654.89802953 & -9.4 & 3.3 \\
        TOI-5205 & 2460907.94012504 & -48.5 & 3.3 & TOI-3714 & 2460654.90985262 & -33.1 & 3.8 \\
        TOI-5205 & 2460907.95173379 & -69.7 & 3.1 & TOI-3714 & 2460654.92086983 & -21.2 & 6.4 \\
        TOI-5205 & 2460907.96333094 & -86.9 & 3.1 & TOI-3714 & 2460654.93289312 & -23.3 & 6.5 \\
        TOI-5205 & 2460907.97492819 & -90.9 & 3.4 & TOI-3714 & 2460654.94497441 & -29.3 & 6.1 \\
        TOI-5205 & 2460907.98653694 & -112.9 & 3.3 & TOI-3714 & 2460654.9562615 & -33.6 & 4.9 \\
        TOI-5205 & 2460920.89851081 & 78.3 & 2.3 & TOI-3714 & 2460654.96796169 & -44.0 & 4.1 \\
        TOI-5205 & 2460920.91010777 & 63.9 & 2.3 & TOI-3714 & 2460654.97926079 & -48.5 & 4.8 \\
        TOI-5205 & 2460920.92170473 & 66.3 & 2.5 & TOI-3714 & 2460654.99127057 & -54.6 & 6.3 \\
        TOI-5205 & 2460920.93331329 & 31.8 & 2.6 & TOI-3714 & 2460934.96477374 & 33.9 & 3.6 \\
        TOI-5205 & 2460920.94491035 & -3.6 & 2.9 & TOI-3714 & 2460934.97638363 & 31.2 & 3.8 \\
        TOI-5205 & 2460920.95650731 & -20.9 & 2.5 & TOI-3714 & 2460934.98798191 & 27.6 & 3.7 \\
        TOI-5205 & 2460920.96811586 & -20.4 & 2.4 & TOI-3714 & 2460934.9995803 & 43.2 & 4.1 \\
        TOI-5205 & 2460920.97971282 & -30.8 & 2.7 & TOI-3714 & 2460935.01119019 & 35.5 & 3.7 \\
        TOI-5205 & 2460920.99132137 & -51.8 & 2.5 & TOI-3714 & 2460935.02278847 & 12.6 & 3.6 \\
        TOI-5205 & 2460921.00291832 & -65.8 & 2.6 & TOI-3714 & 2460935.03439836 & -21.0 & 3.6 \\
        TOI-5205 & 2460921.01452687 & -80.6 & 2.8 & TOI-3714 & 2460935.04599664 & -28.3 & 3.4 \\
        TOI-5205 & 2460921.02612383 & -101.1 & 2.7 & TOI-3714 & 2460935.05759492 & -9.0 & 3.4 \\
        TIC 46432937 & 2461036.90173895 & 172.4 & 2.6 & TOI-3714 & 2460935.06920481 & -16.2 & 3.2 \\
        TIC 46432937 & 2461036.91333594 & 135.2 & 2.6 & TOI-3714 & 2460935.08080309 & -22.3 & 3.4 \\
        TIC 46432937 & 2461036.92494453 & 99.8 & 2.5 & TOI-3714 & 2460935.09240147 & -26.5 & 3.8 \\
        TIC 46432937 & 2461036.93654152 & 65.5 & 2.7 & TOI-3714 & 2460935.10401135 & -34.5 & 3.5 \\
        TIC 46432937 & 2461036.9481501 & 0.2 & 3.0 & & & & \\ \hline
    \end{tabular}
    \label{tab:ap-rvs}
\end{table}

\section{Prior \& Posterior Distributions}\label{ap:posteriors}

The tables in this section list the priors and posterior distributions of the models of the RM effect discussed in Section \ref{subsec:analysis-rvmodel}.


\begin{table}[h]
    \centering
    \caption{Model parameter priors and posterior point estimates for TOI-5205.}
    \begin{tabular}{ccc}
    \hline \hline
    Parameter & Prior & Posterior \\
    \hline
        Stellar mass, $M_\star$ [$M_\odot$] & $\mathcal{N}(0.392, 0.015)^a$ & $0.396_{-0.012}^{+0.013}$ \\
        Stellar radius, $R_\star$ [$R_\odot$]  & $\mathcal{N}(0.394, 0.011)^a$ & $0.386 \pm 0.009$ \\
        Projected stellar rotation velocity, $v \sin i_*$ [km/s] & $\mathcal{U}(0, 2)^a$ & $0.53 \pm 0.04$ \\
        Orbital period, $P$ [days] & $\mathcal{N}(1.630757, 0.000001)^a$ & $1.6307559 \pm 0.0000006$ \\
        Time of mid-transit, $t_0$ [BJD] & $\mathcal{N}(2443.47179, 0.00019)^a$ & $2443.4717 \pm 0.0002$ \\
        RV semi-amplitude, $K$ [m/s] & $\mathcal{N}(346, 14)^a$ & $352 \pm 8$ \\
        Impact parameter, $b$ & $\mathcal{U}(0, 1)^b$ & $0.36 \pm 0.04$ \\
        Planet-to-star radius ratio, $R_p/R_*$ & $\mathcal{SN}(0.2720, 0.0043, 0.0039)^a$ & $0.272 \pm 0.004$ \\
        Projected stellar obliquity $\lambda$ [deg] & $\mathcal{U}(-180, 180)$ & $0_{-5}^{+6}$ \\
        Velocity offset of transit 1, $v_{0,1}$ [m/s] & $\mathcal{U}(-2000, 2000)$ & $21 \pm 3$ \\
        Velocity offset of transit 2, $v_{0,2}$ [m/s] & $\mathcal{U}(-2000, 2000)$ & $14 \pm 1$ \\
        Velocity offset of transit 3, $v_{0,3}$ [m/s] & $\mathcal{U}(-2000, 2000)$ & $23 \pm 1$ \\
        Inclination, $i$ [deg] & --$^c$ & $88.16_{-0.21}^{+0.22}$ \\

    \hline
    \end{tabular}
    
    {{$^a$} Prior obtained from \citet{KanodiaMahadevan2023}.}
    
    {{$^b$} While $b$ was fit, a prior obtained from \citet{KanodiaMahadevan2023} was placed on $i$ directly.}
    
    {{$^c$} Derived from other parameters.}
    
    \label{tab:TOI5205-posterior}
\end{table}

\begin{table}[h]
    \centering
    \caption{Model parameter priors and posterior point estimates for TIC 46432937.}
    \begin{tabular}{ccc}
    \hline \hline
    Parameter & Prior & Posterior \\
    \hline
        Stellar mass, $M_\star$ [$M_\odot$] & $\mathcal{N}(0.563, 0.029)^a$ & $0.577 \pm 0.007$ \\
        Stellar radius, $R_\star$ [$R_\odot$]  & $\mathcal{N}(0.5299, 0.0091)^a$ & $0.528 \pm 0.009$ \\
        Projected stellar rotation velocity, $v \sin i_*$ [km/s] & $\mathcal{U}(0, 2)$ & $1.32_{-0.18}^{+0.20}$ \\
        Orbital period, $P$ [days] & $\mathcal{N}(1.440445270, 0.000000087)^a$ & $1.440445273_{-0.000000087}^{+0.000000088}$ \\
        Time of mid-transit, $t_0$ [BJD] & $\mathcal{N}(2252.288940, 0.000039)^a$ & $2252.28894 \pm 0.00004$ \\
        RV semi-amplitude, $K$ [m/s] & $\mathcal{N}(837.1, 4.4)^a$ & $837 \pm 4$ \\
        Impact parameter, $b$ & $\mathcal{SN}(0.8250, 0.0073, 0.0157)^a$ & $0.83 \pm 0.01$ \\
        Planet-to-star radius ratio, $R_p/R_*$ & $\mathcal{N}(0.2301, 0.0037)^a$ & $0.230 \pm 0.004$ \\
        Projected stellar obliquity $\lambda$ [deg] & $\mathcal{U}(-180, 180)$ & $0 \pm 3$ \\
        Velocity offset, $v_{0}$ [m/s] & $\mathcal{U}(-2000, 2000)$ & $22 \pm 1$ \\
        Inclination, $i$ [deg] & --$^c$ & $84.35 \pm 0.13$ \\

    \hline
    \end{tabular}
    
    {{$^a$} Prior obtained from \citet{HartmanBayliss2024}.}

    {{$^b$} No posterior for $h, k$ were reported in \citet{HartmanBayliss2024}, so only an uninformed prior was put on these parameters.}

    {{$^c$} Derived from other parameters.}

    \label{tab:tic46432937-posterior}
\end{table}

\begin{table}[h]
    \centering
    \caption{Model parameter priors and posterior point estimates for TOI-3714.}
    \begin{tabular}{ccc}
    \hline \hline
    Parameter & Prior & Posterior \\
    \hline
        Stellar mass, $M_\star$ [$M_\odot$] & $\mathcal{N}(0.507, 0.011)^a$ & $0.512_{-0.010}^{+0.011}$ \\
        Stellar radius, $R_\star$ [$R_\odot$] & $\mathcal{N}(0.509, 0.015)^a$ & $0.494 \pm 0.013$ \\
        Projected stellar rotation velocity, $v \sin i_*$ [km/s] & $\mathcal{U}(0,2)^b$ & $1.09_{-0.09}^{+0.10}$ \\
        Orbital period, $P$ [days] & $\mathcal{N}(2.154849, 0.000001)^b$ & $2.154849 8\pm 0.0000007$ \\
        Time of mid-transit, $t_0$ [BJD] & $\mathcal{N}(2458840.5093, 0.0004)^b$ & $2458840.5093 \pm 0.0004$ \\
        RV semi-amplitude, $K$ [m/s] & $\mathcal{SN}(169, 5, 6)^b$ & $176_{-5}^{+4}$ \\
        Impact parameter, $b$ & $\mathcal{SN}(0.26, 0.10, 0.08)^b$ & $0.30_{-0.10}^{+0.08}$ \\
        Planet-to-star radius ratio, $R_p/R_*$ & $\mathcal{N}(0.204, 0.003)^b$ & $0.204 \pm 0.003$ \\
        Projected stellar obliquity $\lambda$ [deg] & $\mathcal{U}(-180, 180)$ & $12_{-7}^{+9}$ \\
        Velocity offset of transit 1, $v_{0,1}$ [m/s] & $\mathcal{U}(-2000, 2000)$ & $0 \pm 1$ \\
        Velocity offset of transit 2, $v_{0,2}$ [m/s] & $\mathcal{U}(-2000, 2000)$ & $7 \pm 1$ \\
        Inclination, $i$ [deg] & --$^c$ & $88.48_{-0.44}^{+0.50}$ \\

    \hline
    \end{tabular}
    
    {{$^a$} Prior obtained from \citetalias{WeissermanGillis2025}.}

    {{$^b$} Prior obtained from \citet{CanasKanodia2022}.}

    {{$^c$} Derived from other parameters.}
    
    \label{tab:TOI3714-posterior}
\end{table}


\clearpage

\bibliography{PASPsample701}{}

@ARTICLE{AmardPalacios2019,
       author = {{Amard}, L. and {Palacios}, A. and {Charbonnel}, C. and {Gallet}, F. and {Georgy}, C. and {Lagarde}, N. and {Siess}, L.},
        title = "{First grids of low-mass stellar models and isochrones with self-consistent treatment of rotation. From 0.2 to 1.5 M$_{☉}$ at seven metallicities from PMS to TAMS}",
      journal = {\aap},
         year = 2019,
        month = nov,
       volume = {631},
          eid = {A77},
        pages = {A77},
          doi = {10.1051/0004-6361/201935160},
archivePrefix = {arXiv},
       eprint = {1905.08516},
 primaryClass = {astro-ph.SR},
       adsurl = {https://ui.adsabs.harvard.edu/abs/2019A&A...631A..77A}
}

@ARTICLE{BeanSeifahrt2010,
       author = {{Bean}, Jacob L. and {Seifahrt}, Andreas and {Hartman}, Henrik and {Nilsson}, Hampus and {Wiedemann}, G{\"u}nter and {Reiners}, Ansgar and {Dreizler}, Stefan and {Henry}, Todd J.},
        title = "{The CRIRES Search for Planets Around the Lowest-mass Stars. I. High-precision Near-infrared Radial Velocities with an Ammonia Gas Cell}",
      journal = {\apj},
         year = 2010,
        month = apr,
       volume = {713},
       number = {1},
        pages = {410-422},
          doi = {10.1088/0004-637X/713/1/410},
archivePrefix = {arXiv},
       eprint = {0911.3148},
 primaryClass = {astro-ph.EP},
       adsurl = {https://ui.adsabs.harvard.edu/abs/2010ApJ...713..410B}
}

@ARTICLE{BehmardDai2022,
       author = {{Behmard}, Aida and {Dai}, Fei and {Howard}, Andrew W.},
        title = "{Stellar Companions to TESS Objects of Interest: A Test of Planet-Companion Alignment}",
      journal = {\aj},
         year = 2022,
        month = apr,
       volume = {163},
       number = {4},
          eid = {160},
        pages = {160},
          doi = {10.3847/1538-3881/ac53a7},
archivePrefix = {arXiv},
       eprint = {2202.01798},
 primaryClass = {astro-ph.EP},
       adsurl = {https://ui.adsabs.harvard.edu/abs/2022AJ....163..160B}
}

@ARTICLE{BeleznayKunimoto2022,
       author = {{Beleznay}, Maya and {Kunimoto}, M.},
        title = "{Exploring the dependence of hot Jupiter occurrence rates on stellar mass with TESS}",
      journal = {\mnras},
         year = 2022,
        month = oct,
       volume = {516},
       number = {1},
        pages = {75-83},
          doi = {10.1093/mnras/stac2179},
archivePrefix = {arXiv},
       eprint = {2207.12522},
 primaryClass = {astro-ph.EP},
       adsurl = {https://ui.adsabs.harvard.edu/abs/2022MNRAS.516...75B}
}

@ARTICLE{BitschSchneider2022,
       author = {{Bitsch}, Bertram and {Schneider}, Aaron David and {Kreidberg}, Laura},
        title = "{How drifting and evaporating pebbles shape giant planets. III. The formation of WASP-77A b and {\ensuremath{\tau}} Bo{\"o}tis b}",
      journal = {\aap},
         year = 2022,
        month = sep,
       volume = {665},
          eid = {A138},
        pages = {A138},
          doi = {10.1051/0004-6361/202243345},
archivePrefix = {arXiv},
       eprint = {2207.06077},
 primaryClass = {astro-ph.EP},
       adsurl = {https://ui.adsabs.harvard.edu/abs/2022A&A...665A.138B}
}

@ARTICLE{Carteret2026,
       author = {{Carteret}, Yann and {Bourrier}, Vincent and {Parc}, L{\'e}na and {Winter}, Andrew and {Allart}, Romain and {de Castro Le{\~a}o}, Izan and {Bouchy}, Fran{\c{c}}ois and {Passegger}, Vera M. and {Cadieux}, Charles and {Lamontagne}, Pierrot and {Artigau}, {\'E}tienne and {Baron}, Fr{\'e}d{\'e}rique and {Barros}, Susana C.~C. and {Benneke}, Bj{\"o}rn and {Bonfils}, Xavier and {Bryan}, Marta and {Canto Martins}, Bruno L. and {Cloutier}, Ryan and {Cristo}, Eduardo and {Gagn{\'e}}, Jonathan and {Cook}, Neil J. and {Cowan}, Nicolas B. and {De Medeiros}, Jose Renan and {Delfosse}, Xavier and {Doyon}, Ren{\'e} and {Dumusque}, Xavier and {Ehrenreich}, David and {Gonz{\'a}lez Hern{\'a}ndez}, Jonay I. and {Lafreni{\`e}re}, David and {Lovis}, Christophe and {Malo}, Lison and {Melo}, Claudio and {Mignon}, Lucile and {Mordasini}, Christoph and {Pepe}, Francesco and {Rebolo}, Rafael and {Rowe}, Jason and {Santos}, Nuno C. and {S{\'e}gransan}, Damien and {Mascare{\~n}o}, Alejandro Su{\'a}rez and {Udry}, St{\'e}phane and {Valencia}, Diana and {Wade}, Gregg and {Al Moulla}, Khaled and {de Souza Barros de Amorim}, Rillck Guilherme and {Dang}, Lisa and {Deibert}, Emily and {Fontinele}, Dasaev O. and {Forveille}, Thierry and {Frensch}, Yolanda G.~C. and {de Lima Gomes}, Roseane and {Mounzer}, Dany and {Pelletier}, Stefan and {Rosener}, Riley and {Skinner}, Bennett Neil and {Srivastava}, Avidaan and {Stefanov}, Atanas K. and {Vaulato}, Valentina and {Wardenier}, Joost P. and {Weisserman}, Drew},
        title = "{Upside down: GJ 3090 b the first retrograde exoplanet around an M dwarf detected with NIRPS}",
      journal = {\aap},
         year = 2026,
        month = sep,
       volume = {713},
          eid = {L15},
        pages = {L15},
          doi = {10.1051/0004-6361/202661989},
archivePrefix = {arXiv},
       eprint = {2609.24870},
 primaryClass = {astro-ph.EP},
       adsurl = {https://ui.adsabs.harvard.edu/abs/2026A&A...713L..15C}
}

@ARTICLE{BourrierZapateroOsorio2022,
       author = {{Bourrier}, V. and {Zapatero Osorio}, M.~R. and {Allart}, R. and {Attia}, M. and {Cretignier}, M. and {Dumusque}, X. and {Lovis}, C. and {Adibekyan}, V. and {Borsa}, F. and {Figueira}, P. and {Gonz{\'a}lez Hern{\'a}ndez}, J.~I. and {Mehner}, A. and {Santos}, N.~C. and {Schmidt}, T. and {Seidel}, J.~V. and {Sozzetti}, A. and {Alibert}, Y. and {Casasayas-Barris}, N. and {Ehrenreich}, D. and {Lo Curto}, G. and {Martins}, C.~J.~A.~P. and {Di Marcantonio}, P. and {M{\'e}gevand}, D. and {Nunes}, N.~J. and {Palle}, E. and {Poretti}, E. and {Sousa}, S.~G.},
        title = "{The polar orbit of the warm Neptune GJ 436b seen with VLT/ESPRESSO}",
      journal = {\aap},
         year = 2022,
        month = jul,
       volume = {663},
          eid = {A160},
        pages = {A160},
          doi = {10.1051/0004-6361/202142559},
archivePrefix = {arXiv},
       eprint = {2203.06109},
 primaryClass = {astro-ph.EP},
       adsurl = {https://ui.adsabs.harvard.edu/abs/2022A&A...663A.160B}
}

@ARTICLE{BoyleBouma2026,
       author = {{Boyle}, Andrew W. and {Bouma}, Luke G. and {Mann}, Andrew W.},
        title = "{The TESS All-Sky Rotation Survey: Periods for 1,046,317 Stars Within 500 pc}",
      journal = {arXiv e-prints},
         year = 2026,
        month = mar,
          eid = {arXiv:2603.05586},
        pages = {arXiv:2603.05586},
          doi = {10.48550/arXiv.2603.05586},
archivePrefix = {arXiv},
       eprint = {2603.05586},
 primaryClass = {astro-ph.SR},
       adsurl = {https://ui.adsabs.harvard.edu/abs/2026arXiv260305586B}
}

@ARTICLE{BryantBayliss2023,
       author = {{Bryant}, Edward M. and {Bayliss}, Daniel and {Van Eylen}, Vincent},
        title = "{The occurrence rate of giant planets orbiting low-mass stars with TESS}",
      journal = {\mnras},
         year = 2023,
        month = may,
       volume = {521},
       number = {3},
        pages = {3663-3681},
          doi = {10.1093/mnras/stad626},
archivePrefix = {arXiv},
       eprint = {2303.00659},
 primaryClass = {astro-ph.EP},
       adsurl = {https://ui.adsabs.harvard.edu/abs/2023MNRAS.521.3663B}
}

@ARTICLE{BryantJordan2025,
       author = {{Bryant}, Edward M. and {Jord{\'a}n}, Andr{\'e}s and {Hartman}, Joel D. and {Bayliss}, Daniel and {Sedaghati}, Elyar and {Barkaoui}, Khalid and {Chouqar}, Jamila and {Pozuelos}, Francisco J. and {Thorngren}, Daniel P. and {Timmermans}, Mathilde and {Almenara}, Jose Manuel and {Chilingarian}, Igor V. and {Collins}, Karen A. and {Gan}, Tianjun and {Howell}, Steve B. and {Narita}, Norio and {Palle}, Enric and {Rackham}, Benjamin V. and {Triaud}, Amaury H.~M.~J. and {Bakos}, Gaspar {\'A}. and {Brahm}, Rafael and {Hobson}, Melissa J. and {Van Eylen}, Vincent and {Amado}, Pedro J. and {Arnold}, Luc and {Bonfils}, Xavier and {Burdanov}, Artem and {Cadieux}, Charles and {Caldwell}, Douglas A. and {Casanova}, Victor and {Charbonneau}, David and {Clark}, Catherine A. and {Collins}, Kevin I. and {Daylan}, Tansu and {Dransfield}, Georgina and {Demory}, Brice-Olivier and {Ducrot}, Elsa and {Fern{\'a}ndez-Rodr{\'\i}guez}, Gareb and {Fukuda}, Izuru and {Fukui}, Akihiko and {Gillon}, Micha{\"e}l and {Gore}, Rebecca and {Hooton}, Matthew J. and {Ikuta}, Kai and {Jehin}, Emmanuel and {Jenkins}, Jon M. and {Levine}, Alan M. and {Littlefield}, Colin and {Murgas}, Felipe and {Nguyen}, Kendra and {Parviainen}, Hannu and {Queloz}, Didier and {Seager}, S. and {Sebastian}, Daniel and {Srdoc}, Gregor and {Vanderspek}, R. and {Winn}, Joshua N. and {de Wit}, Julien and {Z{\'u}{\~n}iga-Fern{\'a}ndez}, Sebasti{\'a}n},
        title = "{A transiting giant planet in orbit around a 0.2-solar-mass host star}",
      journal = {Nature Astronomy},
         year = 2025,
        month = jul,
       volume = {9},
        pages = {1031-1044},
          doi = {10.1038/s41550-025-02552-4},
archivePrefix = {arXiv},
       eprint = {2506.07931},
 primaryClass = {astro-ph.EP},
       adsurl = {https://ui.adsabs.harvard.edu/abs/2025NatAs...9.1031B}
}

@ARTICLE{CanasKanodia2022,
       author = {{Ca{\~n}as}, Caleb I. and {Kanodia}, Shubham and {Bender}, Chad F. and {Mahadevan}, Suvrath and {Stef{\'a}nsson}, Gu{\dj}hmundur and {Cochran}, William D. and {Lin}, Andrea S.~J. and {Hwang}, Hsiang-Chih and {Powers}, Luke and {Monson}, Andrew and {Green}, Elizabeth M. and {Parker}, Brock A. and {Swaby}, Tera N. and {Kobulnicky}, Henry A. and {Wisniewski}, John and {Gupta}, Arvind F. and {Everett}, Mark E. and {Jones}, Sinclaire and {Anjakos}, Benjamin and {Beard}, Corey and {Blake}, Cullen H. and {Diddams}, Scott A. and {Dong}, Zehao and {Fredrick}, Connor and {Hakemiamjad}, Elnaz and {Hebb}, Leslie and {Libby-Roberts}, Jessica E. and {Logsdon}, Sarah E. and {McElwain}, Michael W. and {Metcalf}, Andrew J. and {Ninan}, Joe P. and {Rajagopal}, Jayadev and {Ramsey}, Lawrence W. and {Robertson}, Paul and {Roy}, Arpita and {Ruhle}, Jacob and {Schwab}, Christian and {Terrien}, Ryan C. and {Wright}, Jason T.},
        title = "{TOI-3714 b and TOI-3629 b: Two Gas Giants Transiting M Dwarfs Confirmed with the Habitable-zone Planet Finder and NEID}",
      journal = {\aj},
         year = 2022,
        month = aug,
       volume = {164},
       number = {2},
          eid = {50},
        pages = {50},
          doi = {10.3847/1538-3881/ac7804},
archivePrefix = {arXiv},
       eprint = {2201.09963},
 primaryClass = {astro-ph.EP},
       adsurl = {https://ui.adsabs.harvard.edu/abs/2022AJ....164...50C}
}

@ARTICLE{CanasKanodia2023,
       author = {{Ca{\~n}as}, Caleb I. and {Kanodia}, Shubham and {Libby-Roberts}, Jessica and {Lin}, Andrea S.~J. and {Schutte}, Maria and {Powers}, Luke and {Jones}, Sinclaire and {Monson}, Andrew and {Wang}, Songhu and {Stef{\'a}nsson}, Gu{\dj}mundur and {Cochran}, William D. and {Robertson}, Paul and {Mahadevan}, Suvrath and {Kowalski}, Adam F. and {Wisniewski}, John and {Parker}, Brock A. and {Larsen}, Alexander and {Chapman}, Franklin A.~L. and {Kobulnicky}, Henry A. and {Gupta}, Arvind F. and {Everett}, Mark E. and {Penprase}, Bryan Edward and {Zeimann}, Gregory and {Beard}, Corey and {Bender}, Chad F. and {Col{\'o}n}, Knicole D. and {Diddams}, Scott A. and {Fredrick}, Connor and {Halverson}, Samuel and {Ninan}, Joe P. and {Ramsey}, Lawrence W. and {Roy}, Arpita and {Schwab}, Christian},
        title = "{TOI-3984 A b and TOI-5293 A b: Two Temperate Gas Giants Transiting Mid-M Dwarfs in Wide Binary Systems}",
      journal = {\aj},
         year = 2023,
        month = jul,
       volume = {166},
       number = {1},
          eid = {30},
        pages = {30},
          doi = {10.3847/1538-3881/acdac7},
archivePrefix = {arXiv},
       eprint = {2302.07714},
 primaryClass = {astro-ph.EP},
       adsurl = {https://ui.adsabs.harvard.edu/abs/2023AJ....166...30C}
}

@ARTICLE{ChristianVanderburg2025,
       author = {{Christian}, Sam and {Vanderburg}, Andrew and {Becker}, Juliette and {Kraus}, Adam L. and {Pearce}, Logan and {Collins}, Karen A. and {Rice}, Malena and {Jensen}, Eric L.~N. and {Baker}, David and {Bozza}, Valerio and {Benni}, Paul and {Bieryla}, Allyson and {Binnenfeld}, Avraham and {Collins}, Kevin I. and {Conti}, Dennis M. and {Crossfield}, Ian J.~M. and {Evans}, Phil and {Johnson}, Marshall C. and {Girardin}, Eric and {Gregorio}, Joao and {Lewin}, Pablo and {Mazeh}, Tsevi and {Murgas}, Felipe and {Panahi}, Aviad and {Pozuelos}, Francisco J. and {Radford}, Don J. and {Relles}, Howard M. and {Rodriguez Frustaglia}, Fabian and {Schwarz}, Richard P. and {Srdoc}, Gregor and {Stockdale}, Chris and {Tan}, Thiam-Guan and {Waalkes}, William C. and {Wang}, Gavin and {Wittrock}, Justin and {Zucker}, Shay},
        title = "{Wide Binary Orbits Are Preferentially Aligned with the Orbits of Small Planets, but Probably Not Hot Jupiters}",
      journal = {\aj},
         year = 2025,
        month = jun,
       volume = {169},
       number = {6},
          eid = {308},
        pages = {308},
          doi = {10.3847/1538-3881/adc933},
archivePrefix = {arXiv},
       eprint = {2405.10379},
 primaryClass = {astro-ph.EP},
       adsurl = {https://ui.adsabs.harvard.edu/abs/2025AJ....169..308C}
}

@ARTICLE{EggletonKiseleva-Eggleton2001,
       author = {{Eggleton}, Peter P. and {Kiseleva-Eggleton}, Ludmila},
        title = "{Orbital Evolution in Binary and Triple Stars, with an Application to SS Lacertae}",
      journal = {\apj},
         year = 2001,
        month = dec,
       volume = {562},
       number = {2},
        pages = {1012-1030},
          doi = {10.1086/323843},
archivePrefix = {arXiv},
       eprint = {astro-ph/0104126},
 primaryClass = {astro-ph},
       adsurl = {https://ui.adsabs.harvard.edu/abs/2001ApJ...562.1012E}
}

@ARTICLE{FabryckyTremaine2007,
       author = {{Fabrycky}, Daniel and {Tremaine}, Scott},
        title = "{Shrinking Binary and Planetary Orbits by Kozai Cycles with Tidal Friction}",
      journal = {\apj},
         year = 2007,
        month = nov,
       volume = {669},
       number = {2},
        pages = {1298-1315},
          doi = {10.1086/521702},
archivePrefix = {arXiv},
       eprint = {0705.4285},
 primaryClass = {astro-ph},
       adsurl = {https://ui.adsabs.harvard.edu/abs/2007ApJ...669.1298F}
}

@ARTICLE{FabryckyWinn2009,
       author = {{Fabrycky}, Daniel C. and {Winn}, Joshua N.},
        title = "{Exoplanetary Spin-Orbit Alignment: Results from the Ensemble of Rossiter-McLaughlin Observations}",
      journal = {\apj},
         year = 2009,
        month = may,
       volume = {696},
       number = {2},
        pages = {1230-1240},
          doi = {10.1088/0004-637X/696/2/1230},
archivePrefix = {arXiv},
       eprint = {0902.0737},
 primaryClass = {astro-ph.EP},
       adsurl = {https://ui.adsabs.harvard.edu/abs/2009ApJ...696.1230F}
}

@ARTICLE{ForemanMackeyHogg2013,
       author = {{Foreman-Mackey}, Daniel and {Hogg}, David W. and {Lang}, Dustin and {Goodman}, Jonathan},
        title = "{emcee: The MCMC Hammer}",
      journal = {\pasp},
         year = 2013,
        month = mar,
       volume = {125},
       number = {925},
        pages = {306},
          doi = {10.1086/670067},
archivePrefix = {arXiv},
       eprint = {1202.3665},
 primaryClass = {astro-ph.IM},
       adsurl = {https://ui.adsabs.harvard.edu/abs/2013PASP..125..306F}
}

@ARTICLE{FrenschBouchy2026,
       author = {{Frensch}, Yolanda G.~C. and {Bouchy}, Fran{\c{c}}ois and {Lo Curto}, Gaspare and {L'Heureux}, Alexandrine and {de Lima Gomes}, Roseane and {Faria}, Jo{\~a}o and {Dumusque}, Xavier and {Malo}, Lison and {Cointepas}, Marion and {Srivastava}, Avidaan and {Bonfils}, Xavier and {Delgado-Mena}, Elisa and {Nari}, Nicola and {Artigau}, {\'E}tienne and {Baron}, Fr{\'e}d{\'e}rique and {Barros}, Susana C.~C. and {Benneke}, Bj{\"o}rn and {Bryan}, Marta and {Canto Martins}, Bruno L. and {de Castro Le{\~a}o}, Izan and {Cloutier}, Ryan and {Cook}, Neil J. and {Cowan}, Nicolas B. and {Cristo}, Eduardo and {De Medeiros}, Jose R. and {Delfosse}, Xavier and {Doyon}, Ren{\'e} and {Ehrenreich}, David and {Gonz{\'a}lez Hern{\'a}ndez}, Jonay I. and {Lafreni{\`e}re}, David and {Lovis}, Christophe and {Melo}, Claudio and {Mignon}, Lucile and {Mordasini}, Christoph and {Pepe}, Francesco and {Rebolo}, Rafael and {Rowe}, Jason and {Santos}, Nuno C. and {S{\'e}gransan}, Damien and {Su{\'a}rez Mascare{\~n}o}, Alejandro and {Udry}, St{\'e}phane and {Valencia}, Diana and {Wade}, Gregg and {Al Moulla}, Khaled and {Allart}, Romain and {Almenara}, Jose M. and {Barkaoui}, Khalid and {Cadieux}, Charles and {Castro-Gonz{\'a}lez}, Amadeo and {Collins}, Karen A. and {Fajardo-Acosta}, Sergio B. and {Forveille}, Thierry and {Gan}, Tianjun and {da Silva}, Jo{\~a}o Gomes and {Grieves}, Nolan and {Hobson}, Melissa J. and {Howell}, Steve and {Lamontagne}, Pierrot and {Messamah}, Lina and {Nielsen}, Louise D. and {Osborn}, Ares and {Parc}, L{\'e}na and {Piaulet-Ghorayeb}, Caroline and {Stassun}, Keivan G. and {Stefanov}, Atanas K. and {Striegel}, Stephanie and {Ulmer-Moll}, Sol{\`e}ne and {Vaulato}, Valentina and {Watkins}, Cristilyn N.},
        title = "{TOI-3288 b and TOI-4666 b: Two gas giants transiting low-mass stars characterised by NIRPS}",
      journal = {\aap},
         year = 2026,
        month = mar,
       volume = {707},
          eid = {A73},
        pages = {A73},
          doi = {10.1051/0004-6361/202557656},
archivePrefix = {arXiv},
       eprint = {2510.11703},
 primaryClass = {astro-ph.EP},
       adsurl = {https://ui.adsabs.harvard.edu/abs/2026A&A...707A..73F}
}

@ARTICLE{GanCadieux2023,
       author = {{Gan}, Tianjun and {Cadieux}, Charles and {Jahandar}, Farbod and {Vazan}, Allona and {Wang}, Sharon X. and {Mao}, Shude and {Alvarado-Montes}, Jaime A. and {Lin}, D.~N.~C. and {Artigau}, {\'E}tienne and {Cook}, Neil J. and {Doyon}, Ren{\'e} and {Mann}, Andrew W. and {Stassun}, Keivan G. and {Burgasser}, Adam J. and {Rackham}, Benjamin V. and {Howell}, Steve B. and {Collins}, Karen A. and {Barkaoui}, Khalid and {Shporer}, Avi and {de Leon}, Jerome and {Arnold}, Luc and {Ricker}, George R. and {Vanderspek}, Roland and {Latham}, David W. and {Seager}, Sara and {Winn}, Joshua N. and {Jenkins}, Jon M. and {Burdanov}, Artem and {Charbonneau}, David and {Dransfield}, Georgina and {Fukui}, Akihiko and {Furlan}, Elise and {Gillon}, Micha{\"e}l and {Hooton}, Matthew J. and {Lewis}, Hannah M. and {Littlefield}, Colin and {Mireles}, Ismael and {Narita}, Norio and {Ormel}, Chris W. and {Quinn}, Samuel N. and {Sefako}, Ramotholo and {Timmermans}, Mathilde and {Vezie}, Michael and {de Wit}, Julien},
        title = "{A Massive Hot Jupiter Orbiting a Metal-rich Early M Star Discovered in the TESS Full-frame Images}",
      journal = {\aj},
         year = 2023,
        month = oct,
       volume = {166},
       number = {4},
          eid = {165},
        pages = {165},
          doi = {10.3847/1538-3881/acf56d},
archivePrefix = {arXiv},
       eprint = {2307.07329},
 primaryClass = {astro-ph.EP},
       adsurl = {https://ui.adsabs.harvard.edu/abs/2023AJ....166..165G}
}

@ARTICLE{GanLHeureux2026,
       author = {{Gan}, Tianjun and {L'Heureux}, Alexandrine and {Cadieux}, Charles and {Mao}, Shude and {Pall{\'e}}, Enric and {Wang}, Sharon X. and {Stassun}, Keivan G. and {Howell}, Steve B. and {Rackham}, Benjamin V. and {Grondin}, Steffani M. and {Barkaoui}, Khalid and {Arnold}, Luc and {Artigau}, {\'E}tienne and {Burdanov}, Artem and {Burgasser}, Adam J. and {Caldwell}, Douglas A. and {Ciardi}, David R. and {Collins}, Karen A. and {Cook}, Neil J. and {Doyon}, Ren{\'e} and {Dransfield}, Georgina and {Fukui}, Akihiko and {Gillon}, Micha{\"e}l and {Jehin}, Emmanuel and {Murgas}, Felipe and {Narita}, Norio and {Noguer}, Federico R. and {Relles}, Howard M. and {Shporer}, Avi and {Simon}, Molly N. and {Soubkiou}, Abderahmane and {Theissen}, Christopher A. and {Timmermans}, Mathilde and {Triaud}, Amaury H.~M.~J. and {Zellem}, Robert T. and {Z{\'u}{\~n}iga-Fern{\'a}ndez}, S.},
        title = "{Stellar Multiplicity of M Dwarfs with Short-period Giant Planets, and the Characterization of TOI-5628Ab}",
      journal = {arXiv e-prints},
         year = 2026,
        month = jul,
          eid = {arXiv:2607.10177},
        pages = {arXiv:2607.10177},
archivePrefix = {arXiv},
       eprint = {2607.10177},
 primaryClass = {astro-ph.EP},
       adsurl = {https://ui.adsabs.harvard.edu/abs/2026arXiv260710177G}
}

@ARTICLE{GanWang2023,
       author = {{Gan}, Tianjun and {Wang}, Sharon X. and {Wang}, Songhu and {Mao}, Shude and {Huang}, Chelsea X. and {Collins}, Karen A. and {Stassun}, Keivan G. and {Shporer}, Avi and {Zhu}, Wei and {Ricker}, George R. and {Vanderspek}, Roland and {Latham}, David W. and {Seager}, Sara and {Winn}, Joshua N. and {Jenkins}, Jon M. and {Barkaoui}, Khalid and {Belinski}, Alexander A. and {Ciardi}, David R. and {Evans}, Phil and {Girardin}, Eric and {Maslennikova}, Nataliia A. and {Mazeh}, Tsevi and {Panahi}, Aviad and {Pozuelos}, Francisco J. and {Radford}, Don J. and {Schwarz}, Richard P. and {Twicken}, Joseph D. and {W{\"u}nsche}, Ana{\"e}l and {Zucker}, Shay},
        title = "{Occurrence Rate of Hot Jupiters Around Early-type M Dwarfs Based on Transiting Exoplanet Survey Satellite Data}",
      journal = {\aj},
         year = 2023,
        month = jan,
       volume = {165},
       number = {1},
          eid = {17},
        pages = {17},
          doi = {10.3847/1538-3881/ac9b12},
archivePrefix = {arXiv},
       eprint = {2210.08313},
 primaryClass = {astro-ph.EP},
       adsurl = {https://ui.adsabs.harvard.edu/abs/2023AJ....165...17G}
}

@ARTICLE{GanWang2024,
       author = {{Gan}, Tianjun and {Wang}, Sharon X. and {Dai}, Fei and {Winn}, Joshua N. and {Mao}, Shude and {Xu}, Siyi and {Pall{\'e}}, Enric and {Bean}, Jacob L. and {Brady}, Madison and {Brown}, Nina and {Lu}, Cicero and {Luque}, Rafael and {Mocnik}, Teo and {Seifahrt}, Andreas and {Stef{\'a}nsson}, Gu{\dj}mundur K.},
        title = "{The Aligned Orbit of a Hot Jupiter around the M Dwarf TOI-4201}",
      journal = {\apjl},
         year = 2024,
        month = jul,
       volume = {969},
       number = {1},
          eid = {L24},
        pages = {L24},
          doi = {10.3847/2041-8213/ad5967},
archivePrefix = {arXiv},
       eprint = {2406.12798},
 primaryClass = {astro-ph.EP},
       adsurl = {https://ui.adsabs.harvard.edu/abs/2024ApJ...969L..24G}
}

@ARTICLE{HartmanBakos2023,
       author = {{Hartman}, J.~D. and {Bakos}, G. {\'A}. and {Csubry}, Z. and {Howard}, A.~W. and {Isaacson}, H. and {Giacalone}, S. and {Chontos}, A. and {Narita}, N. and {Fukui}, A. and {de Leon}, J.~P. and {Watanabe}, N. and {Mori}, M. and {Kagetani}, T. and {Fukuda}, I. and {Kawai}, Y. and {Ikoma}, M. and {Palle}, E. and {Murgas}, F. and {Esparza-Borges}, E. and {Parviainen}, H. and {Bouma}, L.~G. and {Cointepas}, M. and {Bonfils}, X. and {Almenara}, J.~M. and {Collins}, Karen A. and {Collins}, Kevin I. and {Relles}, Howard M. and {Barkaoui}, Khalid and {Schwarz}, Richard P. and {Mourad}, Ghachoui and {Timmermans}, Mathilde and {Dransfield}, Georgina and {Burdanov}, Artem and {de Wit}, Julien and {Jehin}, Emmanu{\"e}l and {Triaud}, Amaury H.~M.~J. and {Gillon}, Micha{\"e}l and {Benkhaldoun}, Zouhair and {Horne}, Keith and {Sefako}, Ramotholo and {Jord{\'a}n}, A. and {Brahm}, R. and {Suc}, V. and {Howell}, Steve B. and {Furlan}, E. and {Schlieder}, J.~E. and {Ciardi}, D. and {Barclay}, T. and {Gonzales}, E.~J. and {Crossfield}, I. and {Dressing}, C.~D. and {Goliguzova}, M. and {Tatarnikov}, A. and {Ricker}, George R. and {Vanderspek}, Roland and {Latham}, David W. and {Seager}, S. and {Winn}, Joshua N. and {Jenkins}, Jon M. and {Striegel}, Stephanie and {Shporer}, Avi and {Vanderburg}, Andrew and {Levine}, Alan M. and {Kostov}, Veselin B. and {Watanabe}, David},
        title = "{TOI 4201 b and TOI 5344 b: Discovery of Two Transiting Giant Planets around M-dwarf Stars and Revised Parameters for Three Others}",
      journal = {\aj},
         year = 2023,
        month = oct,
       volume = {166},
       number = {4},
          eid = {163},
        pages = {163},
          doi = {10.3847/1538-3881/acf56e},
archivePrefix = {arXiv},
       eprint = {2307.06809},
 primaryClass = {astro-ph.EP},
       adsurl = {https://ui.adsabs.harvard.edu/abs/2023AJ....166..163H}
}

@ARTICLE{HartmanBayliss2024,
       author = {{Hartman}, Joel D. and {Bayliss}, Daniel and {Brahm}, Rafael and {Bryant}, Edward M. and {Jord{\'a}n}, Andr{\'e}s and {Bakos}, G{\'a}sp{\'a}r {\'A}. and {Hobson}, Melissa J. and {Sedaghati}, Elyar and {Bonfils}, Xavier and {Cointepas}, Marion and {Almenara}, Jose Manuel and {Barkaoui}, Khalid and {Timmermans}, Mathilde and {Dransfield}, George and {Ducrot}, Elsa and {Z{\'u}{\~n}iga-Fern{\'a}ndez}, Sebasti{\'a}n and {Hooton}, Matthew J. and {Pedersen}, Peter Pihlmann and {Pozuelos}, Francisco J. and {Triaud}, Amaury H.~M.~J. and {Gillon}, Micha{\"e}l and {Jehin}, Emmanuel and {Waalkes}, William C. and {Berta-Thompson}, Zachory K. and {Howell}, Steve B. and {Furlan}, Elise and {Ricker}, George R. and {Vanderspek}, Roland and {Seager}, Sara and {Winn}, Joshua N. and {Jenkins}, Jon M. and {Rapetti}, David and {Collins}, Karen A. and {Charbonneau}, David and {Burke}, Christopher J. and {Rodriguez}, David R.},
        title = "{TOI 762 A b and TIC 46432937 b: Two Giant Planets Transiting M-dwarf Stars}",
      journal = {\aj},
         year = 2024,
        month = nov,
       volume = {168},
       number = {5},
          eid = {202},
        pages = {202},
          doi = {10.3847/1538-3881/ad6f07},
archivePrefix = {arXiv},
       eprint = {2407.07187},
 primaryClass = {astro-ph.EP},
       adsurl = {https://ui.adsabs.harvard.edu/abs/2024AJ....168..202H}
}

@ARTICLE{HeBitsch2026,
       author = {{He}, Yaxing and {Bitsch}, Bertram and {Houge}, Adrien and {Williams}, Joe and {Ogihara}, Masahiro},
        title = "{The majority of hot Jupiters formed beyond the water ice line}",
      journal = {arXiv e-prints},
         year = 2026,
        month = jul,
          eid = {arXiv:2607.15144},
        pages = {arXiv:2607.15144},
          doi = {10.48550/arXiv.2607.15144},
archivePrefix = {arXiv},
       eprint = {2607.15144},
 primaryClass = {astro-ph.EP},
       adsurl = {https://ui.adsabs.harvard.edu/abs/2026arXiv260715144H}
}

@ARTICLE{HuangWu2016,
       author = {{Huang}, Chelsea and {Wu}, Yanqin and {Triaud}, Amaury H.~M.~J.},
        title = "{Warm Jupiters Are Less Lonely than Hot Jupiters: Close Neighbors}",
      journal = {\apj},
         year = 2016,
        month = jul,
       volume = {825},
       number = {2},
          eid = {98},
        pages = {98},
          doi = {10.3847/0004-637X/825/2/98},
archivePrefix = {arXiv},
       eprint = {1601.05095},
 primaryClass = {astro-ph.EP},
       adsurl = {https://ui.adsabs.harvard.edu/abs/2016ApJ...825...98H}
}

@ARTICLE{InnanenZheng1997,
       author = {{Innanen}, K.~A. and {Zheng}, J.~Q. and {Mikkola}, S. and {Valtonen}, M.~J.},
        title = "{The Kozai Mechanism and the Stability of Planetary Orbits in Binary Star Systems}",
      journal = {\aj},
         year = 1997,
        month = may,
       volume = {113},
        pages = {1915},
          doi = {10.1086/118405},
       adsurl = {https://ui.adsabs.harvard.edu/abs/1997AJ....113.1915I}
}

@ARTICLE{JohnsonAller2010,
       author = {{Johnson}, John Asher and {Aller}, Kimberly M. and {Howard}, Andrew W. and {Crepp}, Justin R.},
        title = "{Giant Planet Occurrence in the Stellar Mass-Metallicity Plane}",
      journal = {\pasp},
         year = 2010,
        month = aug,
       volume = {122},
       number = {894},
        pages = {905},
          doi = {10.1086/655775},
archivePrefix = {arXiv},
       eprint = {1005.3084},
 primaryClass = {astro-ph.EP},
       adsurl = {https://ui.adsabs.harvard.edu/abs/2010PASP..122..905J}
}

@ARTICLE{KanodiaMahadevan2023,
       author = {{Kanodia}, Shubham and {Mahadevan}, Suvrath and {Libby-Roberts}, Jessica and {Stefansson}, Gudmundur and {Ca{\~n}as}, Caleb I. and {Piette}, Anjali A.~A. and {Boss}, Alan and {Teske}, Johanna and {Chambers}, John and {Zeimann}, Greg and {Monson}, Andrew and {Robertson}, Paul and {Ninan}, Joe P. and {Lin}, Andrea S.~J. and {Bender}, Chad F. and {Cochran}, William D. and {Diddams}, Scott A. and {Gupta}, Arvind F. and {Halverson}, Samuel and {Hawley}, Suzanne and {Kobulnicky}, Henry A. and {Metcalf}, Andrew J. and {Parker}, Brock A. and {Powers}, Luke and {Ramsey}, Lawrence W. and {Roy}, Arpita and {Schwab}, Christian and {Swaby}, Tera N. and {Terrien}, Ryan C. and {Wisniewski}, John},
        title = "{TOI-5205b: A Short-period Jovian Planet Transiting a Mid-M Dwarf}",
      journal = {\aj},
         year = 2023,
        month = mar,
       volume = {165},
       number = {3},
          eid = {120},
        pages = {120},
          doi = {10.3847/1538-3881/acabce},
archivePrefix = {arXiv},
       eprint = {2209.11160},
 primaryClass = {astro-ph.EP},
       adsurl = {https://ui.adsabs.harvard.edu/abs/2023AJ....165..120K}
}

@ARTICLE{LathamRowe2011,
       author = {{Latham}, David W. and {Rowe}, Jason F. and {Quinn}, Samuel N. and {Batalha}, Natalie M. and {Borucki}, William J. and {Brown}, Timothy M. and {Bryson}, Stephen T. and {Buchhave}, Lars A. and {Caldwell}, Douglas A. and {Carter}, Joshua A. and {Christiansen}, Jessie L. and {Ciardi}, David R. and {Cochran}, William D. and {Dunham}, Edward W. and {Fabrycky}, Daniel C. and {Ford}, Eric B. and {Gautier}, III, Thomas N. and {Gilliland}, Ronald L. and {Holman}, Matthew J. and {Howell}, Steve B. and {Ibrahim}, Khadeejah A. and {Isaacson}, Howard and {Jenkins}, Jon M. and {Koch}, David G. and {Lissauer}, Jack J. and {Marcy}, Geoffrey W. and {Quintana}, Elisa V. and {Ragozzine}, Darin and {Sasselov}, Dimitar and {Shporer}, Avi and {Steffen}, Jason H. and {Welsh}, William F. and {Wohler}, Bill},
        title = "{A First Comparison of Kepler Planet Candidates in Single and Multiple Systems}",
      journal = {\apjl},
         year = 2011,
        month = may,
       volume = {732},
       number = {2},
          eid = {L24},
        pages = {L24},
          doi = {10.1088/2041-8205/732/2/L24},
archivePrefix = {arXiv},
       eprint = {1103.3896},
 primaryClass = {astro-ph.EP},
       adsurl = {https://ui.adsabs.harvard.edu/abs/2011ApJ...732L..24L}
}

@ARTICLE{LaughlinBodenheimer2004,
       author = {{Laughlin}, Gregory and {Bodenheimer}, Peter and {Adams}, Fred C.},
        title = "{The Core Accretion Model Predicts Few Jovian-Mass Planets Orbiting Red Dwarfs}",
      journal = {\apjl},
         year = 2004,
        month = sep,
       volume = {612},
       number = {1},
        pages = {L73-L76},
          doi = {10.1086/424384},
archivePrefix = {arXiv},
       eprint = {astro-ph/0407309},
 primaryClass = {astro-ph},
       adsurl = {https://ui.adsabs.harvard.edu/abs/2004ApJ...612L..73L}
}

@ARTICLE{Libby-RobertsSchutte2023,
       author = {{Libby-Roberts}, Jessica E. and {Schutte}, Maria and {Hebb}, Leslie and {Kanodia}, Shubham and {Ca{\~n}as}, Caleb I. and {Stef{\'a}nsson}, Gu{\dh}mundur and {Lin}, Andrea S.~J. and {Mahadevan}, Suvrath and {Parts}, Winter and {Powers}, Luke and {Wisniewski}, John and {Bender}, Chad F. and {Cochran}, William D. and {Diddams}, Scott A. and {Everett}, Mark E. and {Gupta}, Arvind F. and {Halverson}, Samuel and {Kobulnicky}, Henry A. and {Kowalski}, Adam F. and {Larsen}, Alexander and {Monson}, Andrew and {Ninan}, Joe P. and {Parker}, Brock A. and {Ramsey}, Lawrence W. and {Robertson}, Paul and {Schwab}, Christian and {Swaby}, Tera N. and {Terrien}, Ryan C.},
        title = "{An In-depth Look at TOI-3884b: A Super-Neptune Transiting an M4Dwarf with Persistent Starspot Crossings}",
      journal = {\aj},
         year = 2023,
        month = jun,
       volume = {165},
       number = {6},
          eid = {249},
        pages = {249},
          doi = {10.3847/1538-3881/accc2f},
       adsurl = {https://ui.adsabs.harvard.edu/abs/2023AJ....165..249L}
}

@ARTICLE{LugerAgol2019,
       author = {{Luger}, Rodrigo and {Agol}, Eric and {Foreman-Mackey}, Daniel and {Fleming}, David P. and {Lustig-Yaeger}, Jacob and {Deitrick}, Russell},
        title = "{starry: Analytic Occultation Light Curves}",
      journal = {\aj},
         year = 2019,
        month = feb,
       volume = {157},
       number = {2},
          eid = {64},
        pages = {64},
          doi = {10.3847/1538-3881/aae8e5},
archivePrefix = {arXiv},
       eprint = {1810.06559},
 primaryClass = {astro-ph.IM},
       adsurl = {https://ui.adsabs.harvard.edu/abs/2019AJ....157...64L}
}

@ARTICLE{MahBitsch2023,
       author = {{Mah}, Jingyi and {Bitsch}, Bertram and {Pascucci}, Ilaria and {Henning}, Thomas},
        title = "{Close-in ice lines and the super-stellar C/O ratio in discs around very low-mass stars}",
      journal = {\aap},
         year = 2023,
        month = sep,
       volume = {677},
          eid = {L7},
        pages = {L7},
          doi = {10.1051/0004-6361/202347169},
archivePrefix = {arXiv},
       eprint = {2308.15128},
 primaryClass = {astro-ph.EP},
       adsurl = {https://ui.adsabs.harvard.edu/abs/2023A&A...677L...7M}
}

@ARTICLE{MannDupuy2019,
       author = {{Mann}, Andrew W. and {Dupuy}, Trent and {Kraus}, Adam L. and {Gaidos}, Eric and {Ansdell}, Megan and {Ireland}, Michael and {Rizzuto}, Aaron C. and {Hung}, Chao-Ling and {Dittmann}, Jason and {Factor}, Samuel and {Feiden}, Gregory and {Martinez}, Raquel A. and {Ru{\'\i}z-Rodr{\'\i}guez}, Dary and {Thao}, Pa Chia},
        title = "{How to Constrain Your M Dwarf. II. The Mass-Luminosity-Metallicity Relation from 0.075 to 0.70 Solar Masses}",
      journal = {\apj},
         year = 2019,
        month = jan,
       volume = {871},
       number = {1},
          eid = {63},
        pages = {63},
          doi = {10.3847/1538-4357/aaf3bc},
archivePrefix = {arXiv},
       eprint = {1811.06938},
 primaryClass = {astro-ph.SR},
       adsurl = {https://ui.adsabs.harvard.edu/abs/2019ApJ...871...63M}
}

@ARTICLE{MasudaWinn2020,
       author = {{Masuda}, Kento and {Winn}, Joshua N.},
        title = "{On the Inference of a Star's Inclination Angle from its Rotation Velocity and Projected Rotation Velocity}",
      journal = {\aj},
         year = 2020,
        month = mar,
       volume = {159},
       number = {3},
          eid = {81},
        pages = {81},
          doi = {10.3847/1538-3881/ab65be},
archivePrefix = {arXiv},
       eprint = {2001.04973},
 primaryClass = {astro-ph.IM},
       adsurl = {https://ui.adsabs.harvard.edu/abs/2020AJ....159...81M}
}

@ARTICLE{McLaughlin1924,
       author = {{McLaughlin}, D.~B.},
        title = "{Some results of a spectrographic study of the Algol system.}",
      journal = {\apj},
         year = 1924,
        month = jul,
       volume = {60},
        pages = {22-31},
          doi = {10.1086/142826},
       adsurl = {https://ui.adsabs.harvard.edu/abs/1924ApJ....60...22M}
}

@ARTICLE{MolliereMolyarova2022,
       author = {{Molli{\`e}re}, Paul and {Molyarova}, Tamara and {Bitsch}, Bertram and {Henning}, Thomas and {Schneider}, Aaron and {Kreidberg}, Laura and {Eistrup}, Christian and {Burn}, Remo and {Nasedkin}, Evert and {Semenov}, Dmitry and {Mordasini}, Christoph and {Schlecker}, Martin and {Schwarz}, Kamber R. and {Lacour}, Sylvestre and {Nowak}, Mathias and {Schulik}, Matth{\"a}us},
        title = "{Interpreting the Atmospheric Composition of Exoplanets: Sensitivity to Planet Formation Assumptions}",
      journal = {\apj},
         year = 2022,
        month = jul,
       volume = {934},
       number = {1},
          eid = {74},
        pages = {74},
          doi = {10.3847/1538-4357/ac6a56},
archivePrefix = {arXiv},
       eprint = {2204.13714},
 primaryClass = {astro-ph.EP},
       adsurl = {https://ui.adsabs.harvard.edu/abs/2022ApJ...934...74M}
}

@ARTICLE{MoralesMustill2019,
       author = {{Morales}, J.~C. and {Mustill}, A.~J. and {Ribas}, I. and {Davies}, M.~B. and {Reiners}, A. and {Bauer}, F.~F. and {Kossakowski}, D. and {Herrero}, E. and {Rodr{\'\i}guez}, E. and {L{\'o}pez-Gonz{\'a}lez}, M.~J. and {Rodr{\'\i}guez-L{\'o}pez}, C. and {B{\'e}jar}, V.~J.~S. and {Gonz{\'a}lez-Cuesta}, L. and {Luque}, R. and {Pall{\'e}}, E. and {Perger}, M. and {Baroch}, D. and {Johansen}, A. and {Klahr}, H. and {Mordasini}, C. and {Anglada-Escud{\'e}}, G. and {Caballero}, J.~A. and {Cort{\'e}s-Contreras}, M. and {Dreizler}, S. and {Lafarga}, M. and {Nagel}, E. and {Passegger}, V.~M. and {Reffert}, S. and {Rosich}, A. and {Schweitzer}, A. and {Tal-Or}, L. and {Trifonov}, T. and {Zechmeister}, M. and {Quirrenbach}, A. and {Amado}, P.~J. and {Guenther}, E.~W. and {Hagen}, H.-J. and {Henning}, T. and {Jeffers}, S.~V. and {Kaminski}, A. and {K{\"u}rster}, M. and {Montes}, D. and {Seifert}, W. and {Abell{\'a}n}, F.~J. and {Abril}, M. and {Aceituno}, J. and {Aceituno}, F.~J. and {Alonso-Floriano}, F.~J. and {Ammler-von Eiff}, M. and {Antona}, R. and {Arroyo-Torres}, B. and {Azzaro}, M. and {Barrado}, D. and {Becerril-Jarque}, S. and {Ben{\'\i}tez}, D. and {Berdi{\~n}as}, Z.~M. and {Bergond}, G. and {Brinkm{\"o}ller}, M. and {del Burgo}, C. and {Burn}, R. and {Calvo-Ortega}, R. and {Cano}, J. and {C{\'a}rdenas}, M.~C. and {Cardona Guill{\'e}n}, C. and {Carro}, J. and {Casal}, E. and {Casanova}, V. and {Casasayas-Barris}, N. and {Chaturvedi}, P. and {Cifuentes}, C. and {Claret}, A. and {Colom{\'e}}, J. and {Czesla}, S. and {D{\'\i}ez-Alonso}, E. and {Dorda}, R. and {Emsenhuber}, A. and {Fern{\'a}ndez}, M. and {Fern{\'a}ndez-Mart{\'\i}n}, A. and {Ferro}, I.~M. and {Fuhrmeister}, B. and {Galad{\'\i}-Enr{\'\i}quez}, D. and {Gallardo Cava}, I. and {Garc{\'\i}a Vargas}, M.~L. and {Garcia-Piquer}, A. and {Gesa}, L. and {Gonz{\'a}lez-{\'A}lvarez}, E. and {Gonz{\'a}lez Hern{\'a}ndez}, J.~I. and {Gonz{\'a}lez-Peinado}, R. and {Gu{\`a}rdia}, J. and {Guijarro}, A. and {de Guindos}, E. and {Hatzes}, A.~P. and {Hauschildt}, P.~H. and {Hedrosa}, R.~P. and {Hermelo}, I. and {Hern{\'a}ndez Arabi}, R. and {Hern{\'a}ndez Otero}, F. and {Hintz}, D. and {Holgado}, G. and {Huber}, A. and {Huke}, P. and {Johnson}, E.~N. and {de Juan}, E. and {Kehr}, M. and {Kemmer}, J. and {Kim}, M. and {Kl{\"u}ter}, J. and {Klutsch}, A. and {Labarga}, F. and {Labiche}, N. and {Lalitha}, S. and {Lamp{\'o}n}, M. and {Lara}, L.~M. and {Launhardt}, R. and {L{\'a}zaro}, F.~J. and {Lizon}, J.-L. and {Llamas}, M. and {Lodieu}, N. and {L{\'o}pez del Fresno}, M. and {L{\'o}pez Salas}, J.~F. and {L{\'o}pez-Santiago}, J. and {Mag{\'a}n Madinabeitia}, H. and {Mall}, U. and {Mancini}, L. and {Mandel}, H. and {Marfil}, E. and {Mar{\'\i}n Molina}, J.~A. and {Mart{\'\i}n}, E.~L. and {Mart{\'\i}n-Fern{\'a}ndez}, P. and {Mart{\'\i}n-Ruiz}, S. and {Mart{\'\i}nez-Rodr{\'\i}guez}, H. and {Marvin}, C.~J. and {Mirabet}, E. and {Moya}, A. and {Naranjo}, V. and {Nelson}, R.~P. and {Nortmann}, L. and {Nowak}, G. and {Ofir}, A. and {Pascual}, J. and {Pavlov}, A. and {Pedraz}, S. and {P{\'e}rez Medialdea}, D. and {P{\'e}rez-Calpena}, A. and {Perryman}, M.~A.~C. and {Rabaza}, O. and {Ram{\'o}n Ballesta}, A. and {Rebolo}, R. and {Redondo}, P. and {Rix}, H.-W. and {Rodler}, F. and {Rodr{\'\i}guez Trinidad}, A. and {Sabotta}, S. and {Sadegi}, S. and {Salz}, M. and {S{\'a}nchez-Blanco}, E. and {S{\'a}nchez Carrasco}, M.~A. and {S{\'a}nchez-L{\'o}pez}, A. and {Sanz-Forcada}, J. and {Sarkis}, P. and {Sarmiento}, L.~F. and {Sch{\"a}fer}, S. and {Schlecker}, M. and {Schmitt}, J.~H.~M.~M. and {Sch{\"o}fer}, P. and {Solano}, E. and {Sota}, A. and {Stahl}, O. and {Stock}, S. and {Stuber}, T. and {St{\"u}rmer}, J. and {Su{\'a}rez}, J.~C. and {Tabernero}, H.~M. and {Tulloch}, S.~M. and {Veredas}, G. and {Vico-Linares}, J.~I. and {Vilardell}, F. and {Wagner}, K. and {Winkler}, J. and {Wolthoff}, V. and {Yan}, F. and {Zapatero Osorio}, M.~R.},
        title = "{A giant exoplanet orbiting a very-low-mass star challenges planet formation models}",
      journal = {Science},
         year = 2019,
        month = sep,
       volume = {365},
       number = {6460},
        pages = {1441-1445},
          doi = {10.1126/science.aax3198},
archivePrefix = {arXiv},
       eprint = {1909.12174},
 primaryClass = {astro-ph.EP},
       adsurl = {https://ui.adsabs.harvard.edu/abs/2019Sci...365.1441M}
}

@ARTICLE{MustillDavies2015,
       author = {{Mustill}, Alexander J. and {Davies}, Melvyn B. and {Johansen}, Anders},
        title = "{The Destruction of Inner Planetary Systems during High-eccentricity Migration of Gas Giants}",
      journal = {\apj},
         year = 2015,
        month = jul,
       volume = {808},
       number = {1},
          eid = {14},
        pages = {14},
          doi = {10.1088/0004-637X/808/1/14},
archivePrefix = {arXiv},
       eprint = {1502.06971},
 primaryClass = {astro-ph.EP},
       adsurl = {https://ui.adsabs.harvard.edu/abs/2015ApJ...808...14M}
}

@ARTICLE{PascucciHerczeg2013,
       author = {{Pascucci}, I. and {Herczeg}, G. and {Carr}, J.~S. and {Bruderer}, S.},
        title = "{The Atomic and Molecular Content of Disks around Very Low-mass Stars and Brown Dwarfs}",
      journal = {\apj},
         year = 2013,
        month = dec,
       volume = {779},
       number = {2},
          eid = {178},
        pages = {178},
          doi = {10.1088/0004-637X/779/2/178},
archivePrefix = {arXiv},
       eprint = {1311.1228},
 primaryClass = {astro-ph.EP},
       adsurl = {https://ui.adsabs.harvard.edu/abs/2013ApJ...779..178P}
}

@ARTICLE{PenzlinBooth2024,
       author = {{Penzlin}, Anna B.~T. and {Booth}, Richard A. and {Kirk}, James and {Owen}, James E. and {Ahrer}, E. and {Christie}, Duncan A. and {Claringbold}, Alastair B. and {Esparza-Borges}, Emma and {L{\'o}pez-Morales}, M. and {Mayne}, N.~J. and {McCormack}, Mason and {Meech}, Annabella and {Panwar}, Vatsal and {Powell}, Diana and {Sergeev}, Denis E. and {Taylor}, Jake and {Wheatley}, Peter J. and {Zamyatina}, Maria},
        title = "{BOWIE-ALIGN: how formation and migration histories of giant planets impact atmospheric compositions}",
      journal = {\mnras},
         year = 2024,
        month = nov,
       volume = {535},
       number = {1},
        pages = {171-186},
          doi = {10.1093/mnras/stae2362},
archivePrefix = {arXiv},
       eprint = {2407.03199},
 primaryClass = {astro-ph.EP},
       adsurl = {https://ui.adsabs.harvard.edu/abs/2024MNRAS.535..171P}
}

@ARTICLE{RadzomWang2026,
       author = {{Radzom}, Brandon and {Wang}, Songhu and {Pu}, Bonan and {Gautham Bhaskar}, Hareesh and {Rice}, Malena},
        title = "{Hot Jupiters' Isolation Is Not Unique to High-Eccentricity Tidal Migration}",
      journal = {arXiv e-prints},
         year = 2026,
        month = may,
          eid = {arXiv:2605.27362},
        pages = {arXiv:2605.27362},
          doi = {10.48550/arXiv.2605.27362},
archivePrefix = {arXiv},
       eprint = {2605.27362},
 primaryClass = {astro-ph.EP},
       adsurl = {https://ui.adsabs.harvard.edu/abs/2026arXiv260527362R}
}

@ARTICLE{RiceWang2022,
       author = {{Rice}, Malena and {Wang}, Songhu and {Laughlin}, Gregory},
        title = "{Origins of Hot Jupiters from the Stellar Obliquity Distribution}",
      journal = {\apjl},
         year = 2022,
        month = feb,
       volume = {926},
       number = {2},
          eid = {L17},
        pages = {L17},
          doi = {10.3847/2041-8213/ac502d},
archivePrefix = {arXiv},
       eprint = {2201.11768},
 primaryClass = {astro-ph.EP},
       adsurl = {https://ui.adsabs.harvard.edu/abs/2022ApJ...926L..17R}
}

@ARTICLE{RickerWinn2015,
       author = {{Ricker}, George R. and {Winn}, Joshua N. and {Vanderspek}, Roland and {Latham}, David W. and {Bakos}, G{\'a}sp{\'a}r {\'A}. and {Bean}, Jacob L. and {Berta-Thompson}, Zachory K. and {Brown}, Timothy M. and {Buchhave}, Lars and {Butler}, Nathaniel R. and {Butler}, R. Paul and {Chaplin}, William J. and {Charbonneau}, David and {Christensen-Dalsgaard}, J{\o}rgen and {Clampin}, Mark and {Deming}, Drake and {Doty}, John and {De Lee}, Nathan and {Dressing}, Courtney and {Dunham}, Edward W. and {Endl}, Michael and {Fressin}, Francois and {Ge}, Jian and {Henning}, Thomas and {Holman}, Matthew J. and {Howard}, Andrew W. and {Ida}, Shigeru and {Jenkins}, Jon M. and {Jernigan}, Garrett and {Johnson}, John Asher and {Kaltenegger}, Lisa and {Kawai}, Nobuyuki and {Kjeldsen}, Hans and {Laughlin}, Gregory and {Levine}, Alan M. and {Lin}, Douglas and {Lissauer}, Jack J. and {MacQueen}, Phillip and {Marcy}, Geoffrey and {McCullough}, Peter R. and {Morton}, Timothy D. and {Narita}, Norio and {Paegert}, Martin and {Palle}, Enric and {Pepe}, Francesco and {Pepper}, Joshua and {Quirrenbach}, Andreas and {Rinehart}, Stephen A. and {Sasselov}, Dimitar and {Sato}, Bun'ei and {Seager}, Sara and {Sozzetti}, Alessandro and {Stassun}, Keivan G. and {Sullivan}, Peter and {Szentgyorgyi}, Andrew and {Torres}, Guillermo and {Udry}, Stephane and {Villasenor}, Joel},
        title = "{Transiting Exoplanet Survey Satellite (TESS)}",
      journal = {Journal of Astronomical Telescopes, Instruments, and Systems},
         year = 2015,
        month = jan,
       volume = {1},
          eid = {014003},
        pages = {014003},
          doi = {10.1117/1.JATIS.1.1.014003},
archivePrefix = {arXiv},
       eprint = {1406.0151},
 primaryClass = {astro-ph.EP},
       adsurl = {https://ui.adsabs.harvard.edu/abs/2015JATIS...1a4003R}
}

@ARTICLE{Rossiter1924,
       author = {{Rossiter}, R.~A.},
        title = "{On the detection of an effect of rotation during eclipse in the velocity of the brighter component of beta Lyrae, and on the constancy of velocity of this system.}",
      journal = {\apj},
         year = 1924,
        month = jul,
       volume = {60},
        pages = {15-21},
          doi = {10.1086/142825},
       adsurl = {https://ui.adsabs.harvard.edu/abs/1924ApJ....60...15R}
}

@ARTICLE{RusznakWang2025,
       author = {{Rusznak}, Jace and {Wang}, Xian-Yu and {Rice}, Malena and {Wang}, Songhu},
        title = "{From Misaligned Sub-Saturns to Aligned Brown Dwarfs: The Highest M$_{p}$/M$_{*}$ Systems Exhibit Low Obliquities, Even around Hot Stars}",
      journal = {\apjl},
         year = 2025,
        month = apr,
       volume = {983},
       number = {2},
          eid = {L42},
        pages = {L42},
          doi = {10.3847/2041-8213/adc129},
archivePrefix = {arXiv},
       eprint = {2412.04438},
 primaryClass = {astro-ph.EP},
       adsurl = {https://ui.adsabs.harvard.edu/abs/2025ApJ...983L..42R}
}

@ARTICLE{SchmidtSchlaufman2026,
       author = {{Schmidt}, Stephen P. and {Schlaufman}, Kevin C.},
        title = "{Most Hot Jupiters Were Cool Giant Planets for More Than 1 Gyr}",
      journal = {\aj},
         year = 2026,
        month = mar,
       volume = {171},
       number = {3},
          eid = {157},
        pages = {157},
          doi = {10.3847/1538-3881/ae3c11},
archivePrefix = {arXiv},
       eprint = {2601.14367},
 primaryClass = {astro-ph.EP},
       adsurl = {https://ui.adsabs.harvard.edu/abs/2026AJ....171..157S}
}

@INPROCEEDINGS{SeifahrtSturmer2018,
       author = {{Seifahrt}, Andreas and {St{\"u}rmer}, Julian and {Bean}, Jacob L. and {Schwab}, Christian},
        title = "{MAROON-X: a radial velocity spectrograph for the Gemini Observatory}",
    booktitle = {Ground-based and Airborne Instrumentation for Astronomy VII},
         year = 2018,
       editor = {{Evans}, Christopher J. and {Simard}, Luc and {Takami}, Hideki},
       series = {Society of Photo-Optical Instrumentation Engineers (SPIE) Conference Series},
       volume = {10702},
        month = jul,
          eid = {107026D},
        pages = {107026D},
          doi = {10.1117/12.2312936},
archivePrefix = {arXiv},
       eprint = {1805.09276},
 primaryClass = {astro-ph.IM},
       adsurl = {https://ui.adsabs.harvard.edu/abs/2018SPIE10702E..6DS}
}

@INPROCEEDINGS{SeifahrtBean2020,
       author = {{Seifahrt}, Andreas and {Bean}, Jacob L. and {St{\"u}rmer}, Julian and {Kasper}, David and {Gers}, Luke and {Schwab}, Christian and {Zechmeister}, Mathias and {Stef{\'a}nsson}, Gudmundur and {Montet}, Ben and {Dos Santos}, Leonardo A. and {Peck}, Alison and {White}, John and {Tapia}, Eduardo},
        title = "{On-sky commissioning of MAROON-X: a new precision radial velocity spectrograph for Gemini North}",
    booktitle = {Ground-based and Airborne Instrumentation for Astronomy VIII},
         year = 2020,
       editor = {{Evans}, Christopher J. and {Bryant}, Julia J. and {Motohara}, Kentaro},
       series = {Society of Photo-Optical Instrumentation Engineers (SPIE) Conference Series},
       volume = {11447},
        month = dec,
          eid = {114471F},
        pages = {114471F},
          doi = {10.1117/12.2561564},
archivePrefix = {arXiv},
       eprint = {2106.02157},
 primaryClass = {astro-ph.IM},
       adsurl = {https://ui.adsabs.harvard.edu/abs/2020SPIE11447E..1FS}
}

@ARTICLE{SpaldingWinn2022,
       author = {{Spalding}, Christopher and {Winn}, Joshua N.},
        title = "{Tidal Erasure of Stellar Obliquities Constrains the Timing of Hot Jupiter Formation}",
      journal = {\apj},
         year = 2022,
        month = mar,
       volume = {927},
       number = {1},
          eid = {22},
        pages = {22},
          doi = {10.3847/1538-4357/ac4993},
archivePrefix = {arXiv},
       eprint = {2201.03653},
 primaryClass = {astro-ph.EP},
       adsurl = {https://ui.adsabs.harvard.edu/abs/2022ApJ...927...22S}
}

@ARTICLE{TurriniSchisano2021,
       author = {{Turrini}, D. and {Schisano}, E. and {Fonte}, S. and {Molinari}, S. and {Politi}, R. and {Fedele}, D. and {Pani{\'c}}, O. and {Kama}, M. and {Changeat}, Q. and {Tinetti}, G.},
        title = "{Tracing the Formation History of Giant Planets in Protoplanetary Disks with Carbon, Oxygen, Nitrogen, and Sulfur}",
      journal = {\apj},
         year = 2021,
        month = mar,
       volume = {909},
       number = {1},
          eid = {40},
        pages = {40},
          doi = {10.3847/1538-4357/abd6e5},
archivePrefix = {arXiv},
       eprint = {2012.14315},
 primaryClass = {astro-ph.EP},
       adsurl = {https://ui.adsabs.harvard.edu/abs/2021ApJ...909...40T}
}

@ARTICLE{WeissermanGillis2025,
       author = {{Weisserman}, Drew and {Gillis}, Erik and {Cloutier}, Ryan and {Brown}, Nina and {Bean}, Jacob L. and {Seifahrt}, Andreas and {Das}, Tanya and {Brady}, Madison and {Bitsch}, Bertram and {Deibert}, Emily and {Evans-Soma}, Thomas M. and {Fenlon}, Noah and {Kreidberg}, Laura and {Line}, Michael and {Pudritz}, Ralph and {Shkolnik}, Evgenya L. and {Welbanks}, Luis},
        title = "{Aligned Stellar Obliquities for Two Hot Jupiter-hosting M Dwarfs Revealed by MAROON-X: Implications for Hot Jupiter Formation}",
      journal = {\aj},
         year = 2025,
        month = dec,
       volume = {170},
       number = {6},
          eid = {313},
        pages = {313},
          doi = {10.3847/1538-3881/ae08aa},
archivePrefix = {arXiv},
       eprint = {2508.13145},
 primaryClass = {astro-ph.EP},
       adsurl = {https://ui.adsabs.harvard.edu/abs/2025AJ....170..313W}
}

@ARTICLE{WinnFabrycky2010,
       author = {{Winn}, Joshua N. and {Fabrycky}, Daniel and {Albrecht}, Simon and {Johnson}, John Asher},
        title = "{Hot Stars with Hot Jupiters Have High Obliquities}",
      journal = {\apjl},
         year = 2010,
        month = aug,
       volume = {718},
       number = {2},
        pages = {L145-L149},
          doi = {10.1088/2041-8205/718/2/L145},
archivePrefix = {arXiv},
       eprint = {1006.4161},
 primaryClass = {astro-ph.EP},
       adsurl = {https://ui.adsabs.harvard.edu/abs/2010ApJ...718L.145W}
}

@ARTICLE{ZanazziChiang2025,
       author = {{Zanazzi}, J.~J. and {Chiang}, Eugene},
        title = "{Spin and Obliquity Evolution of Hot Jupiter Hosts from Resonance Locks}",
      journal = {\apj},
         year = 2025,
        month = apr,
       volume = {983},
       number = {2},
          eid = {157},
        pages = {157},
          doi = {10.3847/1538-4357/adc114},
archivePrefix = {arXiv},
       eprint = {2410.10943},
 primaryClass = {astro-ph.EP},
       adsurl = {https://ui.adsabs.harvard.edu/abs/2025ApJ...983..157Z}
}

@ARTICLE{Attia2023,
       author = {{Attia}, M. and {Bourrier}, V. and {Delisle}, J.-B. and {Eggenberger}, P.},
        title = "{DREAM: II. The spin─orbit angle distribution of close-in exoplanets under the lens of tides}",
      journal = {\aap},
         year = 2023,
        month = jun,
       volume = {674},
          eid = {A120},
        pages = {A120},
          doi = {10.1051/0004-6361/202245237},
archivePrefix = {arXiv},
       eprint = {2305.00829},
 primaryClass = {astro-ph.EP},
       adsurl = {https://ui.adsabs.harvard.edu/abs/2023A&A...674A.120A}
}

@ARTICLE{ZechmeisterReiners2018,
       author = {{Zechmeister}, M. and {Reiners}, A. and {Amado}, P.~J. and {Azzaro}, M. and {Bauer}, F.~F. and {B{\'e}jar}, V.~J.~S. and {Caballero}, J.~A. and {Guenther}, E.~W. and {Hagen}, H. -J. and {Jeffers}, S.~V. and {Kaminski}, A. and {K{\"u}rster}, M. and {Launhardt}, R. and {Montes}, D. and {Morales}, J.~C. and {Quirrenbach}, A. and {Reffert}, S. and {Ribas}, I. and {Seifert}, W. and {Tal-Or}, L. and {Wolthoff}, V.},
        title = "{Spectrum radial velocity analyser (SERVAL). High-precision radial velocities and two alternative spectral indicators}",
      journal = {\aap},
         year = 2018,
        month = jan,
       volume = {609},
          eid = {A12},
        pages = {A12},
          doi = {10.1051/0004-6361/201731483},
archivePrefix = {arXiv},
       eprint = {1710.10114},
 primaryClass = {astro-ph.IM},
       adsurl = {https://ui.adsabs.harvard.edu/abs/2018A&A...609A..12Z}
}

@ARTICLE{BailerJones2021,
       author = {{Bailer-Jones}, C.~A.~L. and {Rybizki}, J. and {Fouesneau}, M. and {Demleitner}, M. and {Andrae}, R.},
        title = "{Estimating Distances from Parallaxes. V. Geometric and Photogeometric Distances to 1.47 Billion Stars in Gaia Early Data Release 3}",
      journal = {\aj},
         year = 2021,
        month = mar,
       volume = {161},
       number = {3},
          eid = {147},
        pages = {147},
          doi = {10.3847/1538-3881/abd806},
archivePrefix = {arXiv},
       eprint = {2012.05220},
 primaryClass = {astro-ph.SR},
       adsurl = {https://ui.adsabs.harvard.edu/abs/2021AJ....161..147B}
}

@ARTICLE{Allard2001,
       author = {{Allard}, France and {Hauschildt}, Peter H. and {Alexander}, David R. and {Tamanai}, Akemi and {Schweitzer}, Andreas},
        title = "{The Limiting Effects of Dust in Brown Dwarf Model Atmospheres}",
      journal = {\apj},
         year = 2001,
        month = jul,
       volume = {556},
       number = {1},
        pages = {357-372},
          doi = {10.1086/321547},
archivePrefix = {arXiv},
       eprint = {astro-ph/0104256},
 primaryClass = {astro-ph},
       adsurl = {https://ui.adsabs.harvard.edu/abs/2001ApJ...556..357A}
}

@ARTICLE{Stolker2020,
       author = {{Stolker}, T. and {Quanz}, S.~P. and {Todorov}, K.~O. and {K{\"u}hn}, J. and {Molli{\`e}re}, P. and {Meyer}, M.~R. and {Currie}, T. and {Daemgen}, S. and {Lavie}, B.},
        title = "{MIRACLES: atmospheric characterization of directly imaged planets and substellar companions at 4-5 {\ensuremath{\mu}}m. I. Photometric analysis of {\ensuremath{\beta}} Pic b, HIP 65426 b, PZ Tel B, and HD 206893 B}",
      journal = {\aap},
         year = 2020,
        month = mar,
       volume = {635},
          eid = {A182},
        pages = {A182},
          doi = {10.1051/0004-6361/201937159},
archivePrefix = {arXiv},
       eprint = {1912.13316},
 primaryClass = {astro-ph.EP},
       adsurl = {https://ui.adsabs.harvard.edu/abs/2020A&A...635A.182S}
}

@article{Thompson2023,
doi = {10.3847/1538-3881/acf5cc},
url = {https://dx.doi.org/10.3847/1538-3881/acf5cc},
year = {2023},
month = {sep},
publisher = {The American Astronomical Society},
volume = {166},
number = {4},
pages = {164},
author = {William Thompson and Jensen Lawrence and Dori Blakely and Christian Marois and Jason Wang and Mosé Giordano and Timothy Brandt and Doug Johnstone and Jean-Baptiste Ruffio and S. Mark Ammons and Katie A. Crotts and Clarissa R. Do Ó and Eileen C. Gonzales and Malena Rice},
title = {Octofitter: Fast, Flexible, and Accurate Orbit Modeling to Detect Exoplanets},
journal = {The Astronomical Journal},
}

@ARTICLE{Thompson2026,
       author = {{Thompson}, William and {Blakely}, Dori and {Xuan}, Jerry W. and {Blouin}, Simon and {Zhang}, Jingwen and {Johnstone}, Doug and {Ruffio}, Jean-Baptiste and {Nielsen}, Eric and {Speedie}, Jessica and {Bowler}, Brendan P. and {Bouchard-C{\^o}t{\'e}}, Alexandre and {Franson}, Kyle and {Blunt}, Sarah and {Roberson}, William and {Cloutier}, Ryan and {Fogal}, Andre and {Hessel}, Kaitlyn and {Marois}, Christian and {Rochon}, Alexandra},
        title = "{Detecting and Characterizing Companions with a Calibrated Gaia DR2, DR3, and Hipparcos Catalog (G23H)}",
      journal = {arXiv e-prints},
         year = 2026,
        month = jan,
          eid = {arXiv:2602.00235},
        pages = {arXiv:2602.00235},
          doi = {10.48550/arXiv.2602.00235},
archivePrefix = {arXiv},
       eprint = {2602.00235},
 primaryClass = {astro-ph.EP},
       adsurl = {https://ui.adsabs.harvard.edu/abs/2026arXiv260200235T}
}

@article{hgca_dr3,
       author = {Brandt, Timothy D.},
        title = "{The Hipparcos-Gaia Catalog of Accelerations: Gaia {EDR}3 Edition}",
      journal = {\apjs},
         year = 2021,
        month = jun,
       volume = {254},
       number = {2},
        pages = {42},
          doi = {10.3847/1538-4365/abf93c},
       adsurl = {https://ui.adsabs.harvard.edu/abs/2021ApJS..254...42B}
}

@ARTICLE{Kiefer2025,
       author = {{Kiefer}, F. and {Lagrange}, A.-M. and {Rubini}, P. and {Philipot}, F.},
        title = "{Searching for substellar companion candidates with Gaia: I. Introducing the GaiaPMEX tool}",
      journal = {\aap},
         year = 2025,
        month = oct,
       volume = {702},
          eid = {A76},
        pages = {A76},
          doi = {10.1051/0004-6361/202449335},
archivePrefix = {arXiv},
       eprint = {2409.16992},
 primaryClass = {astro-ph.EP},
       adsurl = {https://ui.adsabs.harvard.edu/abs/2025A&A...702A..76K}
}

@ARTICLE{paired,
       author = {{Chance}, Quadry and {Foreman-Mackey}, Daniel and {Ballard}, Sarah and {Casey}, Andrew and {David}, Trevor and {Price-Whelan}, Adrian},
        title = "{paired: A Statistical Framework for Detecting Stellar Binarity with Gaia RVs. I. Sensitivity to Unresolved Binaries}",
      journal = {arXiv e-prints},
         year = 2022,
        month = jun,
          eid = {arXiv:2206.11275},
        pages = {arXiv:2206.11275},
          doi = {10.48550/arXiv.2206.11275},
archivePrefix = {arXiv},
       eprint = {2206.11275},
 primaryClass = {astro-ph.EP},
       adsurl = {https://ui.adsabs.harvard.edu/abs/2022arXiv220611275C}
}

@ARTICLE{Changeat_2025,
       author = {{Changeat}, Quentin and {Lagage}, Pierre-Olivier and {Tinetti}, Giovanna and {Charnay}, Benjamin and {Cowan}, Nicolas B. and {Danielski}, Camilla and {Ducrot}, Elsa and {Dyrek}, Achrene and {Edwards}, Billy and {Ikoma}, Masahiro and {Lueftinger}, Theresa and {Micela}, Giuseppina and {Morello}, Giuseppe and {Panic}, Olja and {Pascale}, Enzo and {Robert}, Severine and {Venot}, Olivia and {Barstow}, Joanna K. and {Bocchieri}, Andrea and {Y-K. Cho}, James and {Cloutier}, Ryan and {Coustenis}, Athena and {Dang}, Lisa and {Fujii}, Yuka and {Ito}, Yuichi and {Lavvas}, Panayotis and {Miguel}, Yamila and {Mugnai}, Lorenzo V. and {Hou Yip}, Kay and {Zak}, Jiri},
        title = "{On the synergetic use of Ariel and JWST for exoplanet atmospheric science}",
      journal = {arXiv e-prints},
         year = 2025,
        month = sep,
          eid = {arXiv:2509.02657},
        pages = {arXiv:2509.02657},
          doi = {10.48550/arXiv.2509.02657},
archivePrefix = {arXiv},
       eprint = {2509.02657},
 primaryClass = {astro-ph.IM},
       adsurl = {https://ui.adsabs.harvard.edu/abs/2025arXiv250902657C}
}

@ARTICLE{WangWang2026,
       author = {{Wang}, Xian-Yu and {Wang}, Songhu and {Ong}, J.~M. Joel},
        title = "{Unified Kraft Break at \raisebox{-0.5ex}\textasciitilde6500 K: A Newly Identified Single-star Obliquity Transition Matches the Classical Rotation Break}",
      journal = {\apjl},
         year = 2026,
        month = jan,
       volume = {996},
       number = {1},
          eid = {L7},
        pages = {L7},
          doi = {10.3847/2041-8213/ae21c5},
archivePrefix = {arXiv},
       eprint = {2511.15610},
 primaryClass = {astro-ph.EP},
       adsurl = {https://ui.adsabs.harvard.edu/abs/2026ApJ...996L...7W}
}

@ARTICLE{RosenthalHoward2024,
       author = {{Rosenthal}, Lee J. and {Howard}, Andrew W. and {Knutson}, Heather A. and {Fulton}, Benjamin J.},
        title = "{The California Legacy Survey. IV. Lonely, Poor, and Eccentric: A Comparison between Solitary and Neighborly Gas Giants}",
      journal = {\apjs},
         year = 2024,
        month = jan,
       volume = {270},
       number = {1},
          eid = {1},
        pages = {1},
          doi = {10.3847/1538-4365/acffc0},
archivePrefix = {arXiv},
       eprint = {2310.08707},
 primaryClass = {astro-ph.EP},
       adsurl = {https://ui.adsabs.harvard.edu/abs/2024ApJS..270....1R}
}

@ARTICLE{FerreiraRice2026,
       author = {{Ferreira}, Thiago and {Rice}, Malena},
        title = "{On the Eccentricity Distribution and Tidal Evolution of Transiting Brown Dwarfs}",
      journal = {\aj},
         year = 2026,
        month = aug,
       volume = {172},
       number = {2},
          eid = {111},
        pages = {111},
          doi = {10.3847/1538-3881/ae8541},
archivePrefix = {arXiv},
       eprint = {2607.02682},
 primaryClass = {astro-ph.EP},
       adsurl = {https://ui.adsabs.harvard.edu/abs/2026AJ....172..111F}
}

@ARTICLE{ShariatElBadry2026,
       author = {{Shariat}, Cheyanne and {El-Badry}, Kareem and {Wang}, Songhu and {Wang}, Xian-Yu and {Rice}, Malena and {Xuan}, Jerry W. and {Ciardi}, David R.},
        title = "{Stellar companions sculpt hot Jupiter formation and spin-orbit evolution}",
      journal = {arXiv e-prints},
         year = 2026,
        month = sep,
          eid = {arXiv:2609.19249},
        pages = {arXiv:2609.19249},
          doi = {10.48550/arXiv.2609.19249},
archivePrefix = {arXiv},
       eprint = {2609.19249},
 primaryClass = {astro-ph.EP},
       adsurl = {https://ui.adsabs.harvard.edu/abs/2026arXiv260919249S}
}

@ARTICLE{WeldonNaoz2024,
       author = {{Weldon}, Grant C. and {Naoz}, Smadar and {Hansen}, Bradley M.~S.},
        title = "{Analytical Models for Secular Descents in Hierarchical Triple Systems}",
      journal = {\apj},
         year = 2024,
        month = oct,
       volume = {974},
       number = {2},
          eid = {302},
        pages = {302},
          doi = {10.3847/1538-4357/ad77a9},
archivePrefix = {arXiv},
       eprint = {2405.20377},
 primaryClass = {astro-ph.EP},
       adsurl = {https://ui.adsabs.harvard.edu/abs/2024ApJ...974..302W}
}

@ARTICLE{RiceWangWang2022,
       author = {{Rice}, Malena and {Wang}, Songhu and {Wang}, Xian-Yu and {Stef{\'a}nsson}, Gu{\dj}mundur and {Isaacson}, Howard and {Howard}, Andrew W. and {Logsdon}, Sarah E. and {Schweiker}, Heidi and {Dai}, Fei and {Brinkman}, Casey and et al.},
        title = "{A Tendency Toward Alignment in Single-star Warm-Jupiter Systems}",
      journal = {\aj},
         year = 2022,
        month = sep,
       volume = {164},
       number = {3},
          eid = {104},
        pages = {104},
          doi = {10.3847/1538-3881/ac8153},
archivePrefix = {arXiv},
       eprint = {2207.06511},
 primaryClass = {astro-ph.EP},
       adsurl = {https://ui.adsabs.harvard.edu/abs/2022AJ....164..104R}
}

@ARTICLE{Zahn1977,
       author = {{Zahn}, J.-P.},
        title = "{Tidal friction in close binary systems.}",
      journal = {\aap},
         year = 1977,
        month = may,
       volume = {57},
        pages = {383-394},
       adsurl = {https://ui.adsabs.harvard.edu/abs/1977A&A....57..383Z}
}

@ARTICLE{BaraffeHomeier2015,
       author = {{Baraffe}, Isabelle and {Homeier}, Derek and {Allard}, France and {Chabrier}, Gilles},
        title = "{New evolutionary models for pre-main sequence and main sequence low-mass stars down to the hydrogen-burning limit}",
      journal = {\aap},
         year = 2015,
        month = may,
       volume = {577},
          eid = {A42},
        pages = {A42},
          doi = {10.1051/0004-6361/201425481},
archivePrefix = {arXiv},
       eprint = {1503.04107},
 primaryClass = {astro-ph.SR},
       adsurl = {https://ui.adsabs.harvard.edu/abs/2015A&A...577A..42B}
}

@ARTICLE{AlbrechtDawson2022,
       author = {{Albrecht}, Simon H. and {Dawson}, Rebekah I. and {Winn}, Joshua N.},
        title = "{Stellar Obliquities in Exoplanetary Systems}",
      journal = {\pasp},
         year = 2022,
        month = aug,
       volume = {134},
       number = {1038},
          eid = {082001},
        pages = {082001},
          doi = {10.1088/1538-3873/ac6c09},
archivePrefix = {arXiv},
       eprint = {2203.05460},
 primaryClass = {astro-ph.EP},
       adsurl = {https://ui.adsabs.harvard.edu/abs/2022PASP..134h2001A}
}

@ARTICLE{WagnerApai2015,
       author = {{Wagner}, Kevin and {Apai}, Daniel and {Kasper}, Markus and {Robberto}, Massimo},
        title = "{Discovery of a Two-armed Spiral Structure in the Gapped Disk around Herbig Ae Star HD 100453}",
      journal = {\apjl},
         year = 2015,
        month = nov,
       volume = {813},
       number = {1},
          eid = {L2},
        pages = {L2},
          doi = {10.1088/2041-8205/813/1/L2},
archivePrefix = {arXiv},
       eprint = {1510.02212},
 primaryClass = {astro-ph.EP},
       adsurl = {https://ui.adsabs.harvard.edu/abs/2015ApJ...813L...2W}
}

@ARTICLE{ReynoldsTobin2021,
       author = {{Reynolds}, Nickalas K. and {Tobin}, John J. and {Sheehan}, Patrick and {Sadavoy}, Sarah I. and {Kratter}, Kaitlin M. and {Li}, Zhi-Yun and {Chandler}, Claire J. and {Segura-Cox}, Dominique and {Looney}, Leslie W. and {Dunham}, Michael M.},
        title = "{Kinematic Analysis of a Protostellar Multiple System: Measuring the Protostar Masses and Assessing Gravitational Instability in the Disks of L1448 IRS3B and L1448 IRS3A}",
      journal = {\apjl},
         year = 2021,
        month = jan,
       volume = {907},
       number = {1},
          eid = {L10},
        pages = {L10},
          doi = {10.3847/2041-8213/abcc02},
archivePrefix = {arXiv},
       eprint = {2011.08293},
 primaryClass = {astro-ph.GA},
       adsurl = {https://ui.adsabs.harvard.edu/abs/2021ApJ...907L..10R}
}

@ARTICLE{Lai2012,
       author = {{Lai}, Dong},
        title = "{Tidal dissipation in planet-hosting stars: damping of spin-orbit misalignment and survival of hot Jupiters}",
      journal = {\mnras},
         year = 2012,
        month = jun,
       volume = {423},
       number = {1},
        pages = {486-492},
          doi = {10.1111/j.1365-2966.2012.20893.x},
archivePrefix = {arXiv},
       eprint = {1109.4703},
 primaryClass = {astro-ph.EP},
       adsurl = {https://ui.adsabs.harvard.edu/abs/2012MNRAS.423..486L}
}

@ARTICLE{LinOgilvie2017,
       author = {{Lin}, Yufeng and {Ogilvie}, Gordon I.},
        title = "{Tidal interactions in spin-orbit misaligned systems}",
      journal = {\mnras},
         year = 2017,
        month = jun,
       volume = {468},
       number = {2},
        pages = {1387-1397},
          doi = {10.1093/mnras/stx540},
archivePrefix = {arXiv},
       eprint = {1702.08805},
 primaryClass = {astro-ph.EP},
       adsurl = {https://ui.adsabs.harvard.edu/abs/2017MNRAS.468.1387L}
}

@ARTICLE{PenevBouma2018,
       author = {{Penev}, Kaloyan and {Bouma}, L.~G. and {Winn}, Joshua N. and {Hartman}, Joel D.},
        title = "{Empirical Tidal Dissipation in Exoplanet Hosts From Tidal Spin-up}",
      journal = {\aj},
         year = 2018,
        month = apr,
       volume = {155},
       number = {4},
          eid = {165},
        pages = {165},
          doi = {10.3847/1538-3881/aaaf71},
archivePrefix = {arXiv},
       eprint = {1802.05269},
 primaryClass = {astro-ph.SR},
       adsurl = {https://ui.adsabs.harvard.edu/abs/2018AJ....155..165P}
}

@ARTICLE{AndersonWinn2021,
       author = {{Anderson}, Kassandra R. and {Winn}, Joshua N. and {Penev}, Kaloyan},
        title = "{On a Possible Solution to the Tidal Realignment Problem for Hot Jupiters}",
      journal = {\apj},
         year = 2021,
        month = jun,
       volume = {914},
       number = {1},
          eid = {56},
        pages = {56},
          doi = {10.3847/1538-4357/abf8af},
archivePrefix = {arXiv},
       eprint = {2102.01081},
 primaryClass = {astro-ph.EP},
       adsurl = {https://ui.adsabs.harvard.edu/abs/2021ApJ...914...56A}
}

@ARTICLE{MessamahBouchy2026,
       author = {{Messamah}, L.~Y. and {Bouchy}, F. and {Venturini}, J. and {Parc}, L. and {Nigioni}, A.},
        title = "{Mind the companion: Demographics of transiting S-type exoplanets}",
      journal = {\aap},
         year = 2026,
        month = aug,
       volume = {712},
          eid = {A214},
        pages = {A214},
          doi = {10.1051/0004-6361/202659086},
archivePrefix = {arXiv},
       eprint = {2605.31523},
 primaryClass = {astro-ph.EP},
       adsurl = {https://ui.adsabs.harvard.edu/abs/2026A&A...712A.214M}
}

@ARTICLE{Naoz2016,
       author = {{Naoz}, Smadar},
        title = "{The Eccentric Kozai-Lidov Effect and Its Applications}",
      journal = {\araa},
         year = 2016,
        month = sep,
       volume = {54},
        pages = {441-489},
          doi = {10.1146/annurev-astro-081915-023315},
archivePrefix = {arXiv},
       eprint = {1601.07175},
 primaryClass = {astro-ph.EP},
       adsurl = {https://scixplorer.org/abs/2016ARA&A..54..441N}
}

@ARTICLE{NgoKnutson2015,
       author = {{Ngo}, Henry and {Knutson}, Heather A. and {Hinkley}, Sasha and {Crepp}, Justin R. and {Bechter}, Eric B. and {Batygin}, Konstantin and {Howard}, Andrew W. and {Johnson}, John A. and {Morton}, Timothy D. and {Muirhead}, Philip S.},
        title = "{Friends of Hot Jupiters. II. No Correspondence between Hot-jupiter Spin-Orbit Misalignment and the Incidence of Directly Imaged Stellar Companions}",
      journal = {\apj},
         year = 2015,
        month = feb,
       volume = {800},
       number = {2},
          eid = {138},
        pages = {138},
          doi = {10.1088/0004-637X/800/2/138},
archivePrefix = {arXiv},
       eprint = {1501.00013},
 primaryClass = {astro-ph.EP},
       adsurl = {https://scixplorer.org/abs/2015ApJ...800..138N}
}

@ARTICLE{ZanazziDewberry2024,
       author = {{Zanazzi}, J.~J. and {Dewberry}, Janosz and {Chiang}, Eugene},
        title = "{Damping Obliquities of Hot Jupiter Hosts by Resonance Locking}",
      journal = {\apjl},
         year = 2024,
        month = jun,
       volume = {967},
       number = {2},
          eid = {L29},
        pages = {L29},
          doi = {10.3847/2041-8213/ad4644},
archivePrefix = {arXiv},
       eprint = {2403.05616},
 primaryClass = {astro-ph.EP},
       adsurl = {https://scixplorer.org/abs/2024ApJ...967L..29Z}
}
\bibliographystyle{aasjournalv7}



\end{document}